\documentclass[reprint,superscriptaddress,amsmath,amssymb,aps]{revtex4-2}

\usepackage{graphicx}
\usepackage{dcolumn}
\usepackage{bm}
\usepackage{ulem}
\usepackage{bm}
\usepackage{algorithm}
\usepackage[noend]{algpseudocode} 
\usepackage{amssymb}
\usepackage{soul}
\usepackage[mathlines]{lineno}
\usepackage{tabularx}
\usepackage{graphicx}
\usepackage{dcolumn}
\usepackage{bm}
\usepackage{amsmath}
\usepackage{xcolor} 
\usepackage[pagebackref]{hyperref} 
\hypersetup{
	colorlinks   = true, 
	urlcolor     = blue, 
	linkcolor    = blue, 
	citecolor   = blue 
}
\usepackage{graphicx}
\usepackage{physics}
\usepackage{mathtools}
\usepackage{dsfont}
\usepackage{natbib}

\begin{document}
\title{
Quantum skyrmion parallelism via metasurface-tailored high-dimensional entanglement}

\author{Pedro Ornelas}
\address{School of Physics, University of the Witwatersrand, Private Bag 3, Wits 2050, South Africa}

\author{Ramona Bedford}
\address{School of Physics and Astronomy, Monash University, Melbourne, Victoria, Australia}

\author{Fazilah Nothlawala}
\address{School of Physics, University of the Witwatersrand, Private Bag 3, Wits 2050, South Africa}

\author{Chi Li}
\address{School of Physics and Astronomy, Monash University, Melbourne, Victoria, Australia}

\author{Haoyi Yu}
\address{School of Physics and Astronomy, Monash University, Melbourne, Victoria, Australia}

\author{Isaac Nape}
\address{School of Physics, University of the Witwatersrand, Private Bag 3, Wits 2050, South Africa}

\author{Stefan A. Maier}
\email{stefan.maier@monash.edu}
\address{School of Physics and Astronomy, Monash University, Melbourne, Victoria, Australia}
\address{Department of Physics, Imperial College London, London, UK}

\author{Haoran Ren}
\email{haoran.ren@monash.edu}
\address{School of Physics and Astronomy, Monash University, Melbourne, Victoria, Australia}

\author{Andrew Forbes}
\email{andrew.forbes@wits.ac.za}
\address{School of Physics, University of the Witwatersrand, Private Bag 3, Wits 2050, South Africa}

\vspace{10pt}

\begin{abstract}
\noindent \textbf{Quantum skyrmions are topological structures that have garnered significant interest due to their demonstrated robustness and versatility across diverse optical platforms. However, existing approaches for their generation are limited to producing pre-determined two dimensional qubit states with a single topology. Here we create multi-dimensional topological states by introducing a non-local interaction between high-dimensional photonic entanglement and a metasurface, where the topological transformation induced by the metasurface is made non-deterministic by the  
probabilistic nature of the interfacing entangled state. Within this framework, we demonstrate that individual quantum states can host multiple co-existing topologies that are only revealed upon measurement, allowing for their parallel transport within distinct spatial mode channels. We confirm this by revealing the rich topological landscape within our modified Hilbert space while controlling the desired output topology by orbital angular momentum (OAM) projections on one of the entangled photons, producing multiple non-local polarization-OAM entangled states characterized by distinct topological classes, all from a single metasurface device. Our results reveal new capability when structured high-dimensional entanglement is interfaced with structured matter capable of coupling photonic degrees of freedom, establishing a new pathway for the compact generation of complex quantum states.}
\end{abstract}

\vspace{2pc}
\maketitle


\noindent In recent years, topologically non-trivial field configurations known as optical Skyrmions have garnered significant interest 
emerging across a wide range of photonic platforms \cite{shen2024optical, yang2025optical, lei2025topological}. 
Their demonstrated topological resilience across diverse noisy environments \cite{wang2024topological, wang2025topological, guo2025topological, peters2026topological, peters2025seeing, ornelas2025topological, de2025quantum, kleine2026topological} has made them desirable candidates for high capacity information storage \cite{wang2025generation, zeng2025tailoring} optical communication and computing \cite{wang2025perturbation}. Notably, it has been shown that the correlations between the spin and orbital angular momentum (OAM) degrees of freedom of two entangled photons can form non-local topological structures \cite{ornelas2024non, koni2025dual}, opening up potential new avenues for the distribution of topology across vast distances.\\
 
\noindent Furthermore, advances in metasurface and metamaterial designs \cite{xie2026metasurfaces, lei2025topological} have enabled for the compact generation of diverse classical and quantum optical Skyrmions realized as near-field tunable surface plasmon \cite{tsesses2018evanescent, schwab2025skyrmion} and phonon polaritons \cite{bau2026tunable, mangold2026phonon}, single-photon \cite{liu2026chip} and classical \cite{lin2024chip} Skyrmions generated from on-chip devices and nanophotonic realizations emerging from semiconductor cavity quantum electrodynamics systems  
\cite{ma2025nanophotonic, ma2025bright}.
In particular, advances in metasurface designs enabling arbitrary coupling between spin and OAM degrees of freedom (DoFs) \cite{bouchard2014optical, devlin2017arbitrary, balthasar2017metasurface} have facilitated the generation of optical beams with spatially-varying polarization (vector beams) \cite{yue2016vector, bao2020minimalist}, and hybrid OAM-spin entangled states \cite{stav2018quantum} within compact and modular experimental configurations. More recently, these developments have enabled the realization of diverse classical Stokes Skyrmion structures \cite{li2024realization} alongside improved designs supporting topological invariance under propagation \cite{mata2025tailoring}, topological tunability during propagation \cite{vogliardioptical}, implementation in metafibres exhibiting subwavelength polarization features \cite{he2024optical} and even enabling the formation of Skyrmionic structures at the focal plane of meta-lenses from scalar beams \cite{mata2025skyrmionic}. 
Despite these advances, the use of metasurfaces in the quantum regime, particularly for the generation of non-local skyrmionic states 
remains unexplored. More importantly, the realization of optical Skyrmions using compact metasurface-based platforms remains constrained by limited capacity 
as states are typically generated with a single topological configuration at a time.\\ 

\begin{figure*}[t]
\centering\includegraphics[width=1\linewidth]{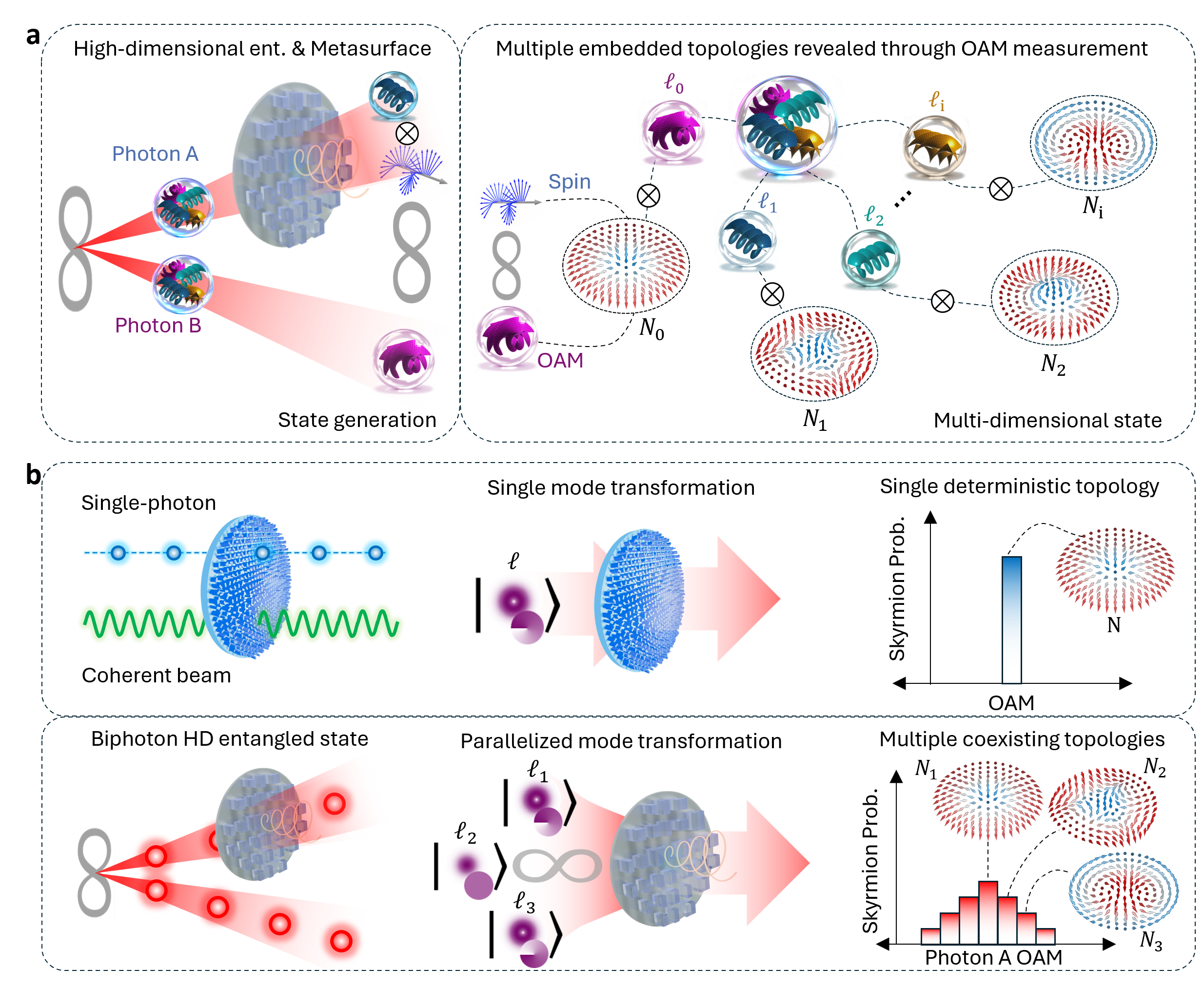}
	\caption{\textbf{Multi-channel quantum topology generated from metasurface and high-dimensional entangled OAM state.} \textbf{a} By interfacing high-dimensional biphoton entanglement with a metasurface, multi-dimensional states are generated with several embedded topologies, derived from the joint correlations of the polarization of photon A and OAM of photon B, and only revealed through OAM measurements on photon A. 
    \textbf{b} Single photons and coherent beams incident on a metasurface yield single mode transformations with a single deterministic topological structure. In contrast, biphoton high-dimensional (HD) entangled states incident on a metasurface allows for multiple mode transformations occurring in parallel, thereby resulting in a complex final topological state possessing many coexisting topological structures.}
	\label{fig: Concept}
\end{figure*}

\noindent Here we propose a scheme to enhance the generation of Skyrmionic states by combining the parallelism offered by high-dimensional biphoton entanglement with a metasurface. 
By sending a single photon of a high-dimensional OAM–OAM entangled state through a metasurface coupling polarization and OAM, as illustrated in Fig.~\ref{fig: Concept} \textbf{a}, a multi-dimensional state is produced
capable of 
hosting a variety of quantum Skyrmions. Therefore 
a wide range of quantum Skyrmion states emerge as a shared property of two photons and are generated using a single passive metasurface. Within this framework, distinct topological states are coupled to the OAM of one photon giving rise to several coexisting quantum topologies embedded within the correlations of a single quantum state, only revealed upon an OAM measurement on one of the photons. We experimentally validate this scheme by revealing the nature of the correlations within the modified Hilbert space after passing an initial high-dimensional state through different metasurfaces. While controlling the desired output state by OAM projections on one of the entangled photons, we demonstrate the creation of multiple topological states from a single metasurface device, 
measuring up to three distinct topological states by heralded OAM measurements. 
We further extend this paradigm to show the generation of diverse and complex topological states, through the projection onto OAM superposition states, emphasising the role of the high-dimensional state enhancement. Our results reveal a new capability when structured entangled photons are interfaced with structured matter, allowing for the direct enhancement of topological information capacity beyond what is capable with single photons or single coherent beams. 
This work establishes a new pathway for the compact generation of complex states that can play host to multi-channel topological networks and supports their integration into miniaturized platforms for robust, high-dimensional quantum communication and information processing.

\section*{Parallelized topological state production with metasurface tailored high-dimensional entanglement}

\noindent Fig.~\ref{fig: Concept}\textbf{b} summarizes the key differences between our proposed scheme and the typical optical Skyrmion generation paradigm. A typical free-space optical Stokes Skyrmion can be generated by passing a single photon or coherent beam occupying any given OAM state 
through a device, such as a J-plate metasurface, which couples its internal polarization and OAM DoFs. The resultant polarization-OAM coupled state is generated as a result of a single mode transformation. The final state then admits a map between real space, $R^2$, and the Poincar\'e sphere, $S^2$, which is represented by the vector field $\vec{S}(\vec{r})$ and characterized by the topological wrapping (Skyrmion) number, $N$, which counts the number of times $S^2$ is wrapped when traversing $R^2$ once. 
Here, the design of the metasurface along with the chosen input OAM state determines the resultant topological structure. However, this scheme limits the generation of topology to a singular topological structure at a time.  
This paradigm can be broken by leveraging the parallelism offered by high-dimensional entanglement. Our scheme involves preparing a high-dimensional OAM-OAM biphoton (photon A and B) entangled state 
with photon A passed through the metasurface as an illustrative example. As before, this results in the coupling of photon A's internal polarization and OAM DoFs. It has been shown that such a transformation results in the generation of hybrid polarization-OAM states with a well-defined topology derived from the correlations of the joint state \cite{ornelas2024non, ornelas2026reconfigurable}. 
Some exemplar topologies are represented in the final panel of Fig.~\ref{fig: Concept}\textbf{b} by their corresponding quantum Stokes vector field, $\vec{S}_A(\vec{r}_B)$, which summarizes the correlation structure of a given hybrid entangled state. 
This vector field encodes the relationship between a given spatial measurement (vector position) on photon B and the corresponding polarization state (vector orientation) on photon A. The topology of the state is then characterized by the number of times the space of photon B wraps that of photon A, i.e., the number of  times the vector field $\vec{S}_A(\vec{r}_B)$ traces out the space of photon A given that we traverse all of the space (spatial positions, $\vec{r}_B$) for photon B. 
Importantly, since the OAM of photon B is unknown until an OAM measurement is performed on photon A, the transformation performed by the metasurface is enacted over all input OAM modes for photon B (or equivalently for photon A) in parallel. In this way the resultant state is not a single topological state, but rather several topological states, each correlated with a distinct OAM state, represented in the final panel of Fig.~\ref{fig: Concept}\textbf{b} as a probability distribution over multiple possible topologies as a function of the chosen OAM measurement of photon A.
The use of a single metasurface enables the passive and parallel generation of multiple distinct spatial channels (in photon A), each characterised by a unique hybrid state and an associated topological number (shared between photon A and B), by shifting the complexity away from the state-generation optics and only to the post-selection optics for the spatial degree of freedom of photon A -- a key feature that distinguishes this work. Next, we outline the protocol for generating our multi-dimensional Skyrmion states from a higher dimensional entangled state. 

\section*{Metasurface design and high-dimensional state enhancement}
\begin{figure*}[t]
\centering\includegraphics[width=1\linewidth]{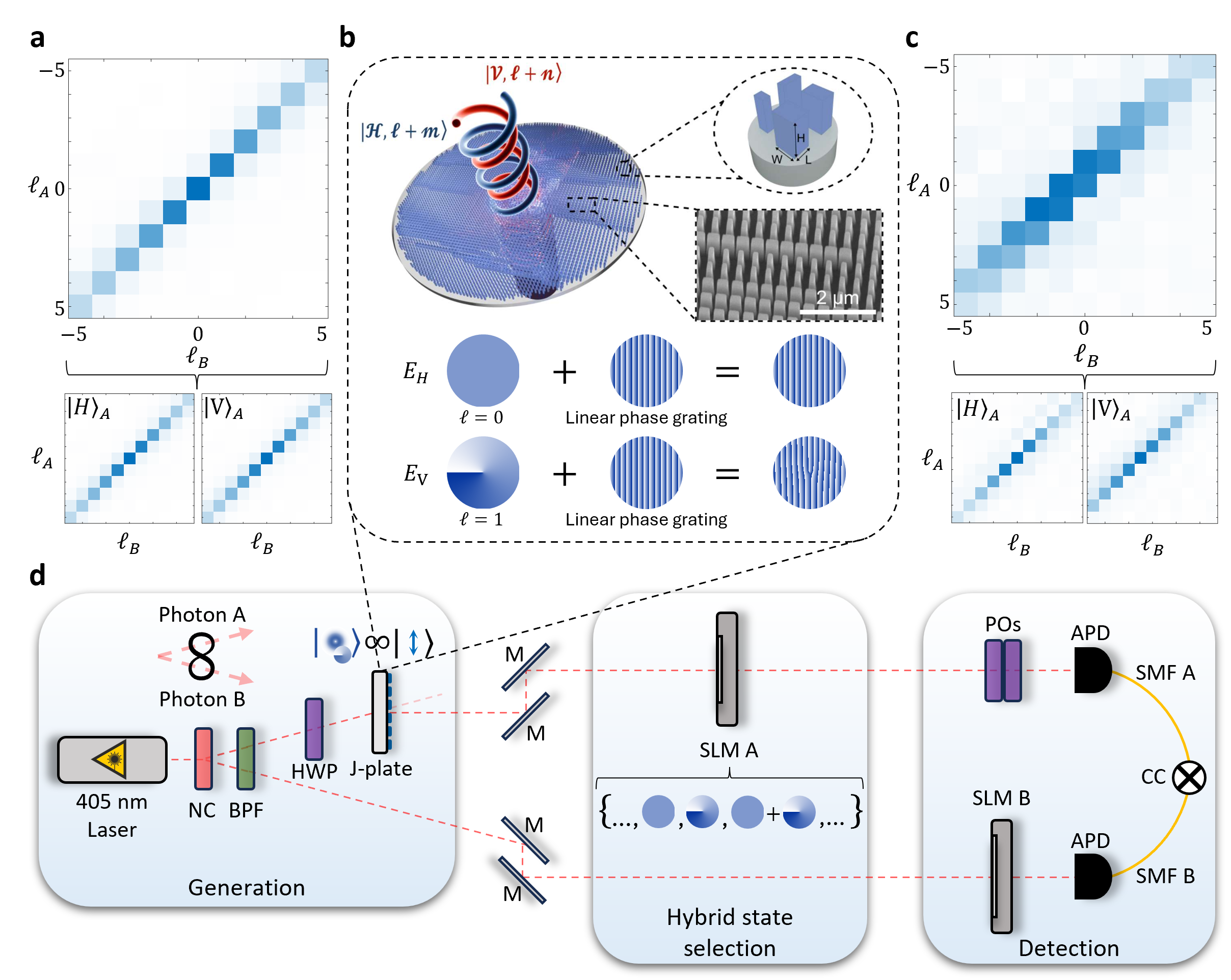}
	\caption{\textbf{Metasurface design and characterization.} \textbf{a}, OAM spiral bandwidth depicting the nature of the high-dimensional state correlations produced directly after the SPDC process (after having passed through the half-wave plate (HWP)). \textbf{b}, Schematic of the metasurface generating spatially overlapped beams with independent phase control in decoupled polarization channels. Insets show the unit-cell geometry, consisting of amorphous silicon (a-Si) cuboid nanopillars. The lower panel illustrates the azimuthal OAM phase profiles with an additional linear phase gradient encoded for both channels. \textbf{c}, Shifted OAM spiral bandwidth measured after passing the OAM-OAM entangled state through the J-plate metasurface. \textbf{d}, Experimental setup for the generation and detection of hybrid OAM-spin entangled states. The generation scheme consists of a typical spontaneous-parameteric down conversion (SPDC) process and spin-polarization coupling device, J-plate. An SLM placed in the path of photon A, post-selects the hybrid subspace to be measured through the projection of photon A's OAM.}
	\label{fig: MetaCharacter}
\end{figure*}

\noindent The experimental setup and metasurface design used to generate the desired multi-dimensional hybrid entangled state is shown in Fig.~\ref{fig: MetaCharacter}. An initial high-dimensional OAM-OAM entangled pair of photons is produced via a spontaneous parametric down-conversion (SPDC) process by sending a pump beam with a Gaussian profile through a non-linear crystal. The resultant biphoton entangled state produced from this process can be expressed as  
\begin{equation}
    \ket{\Psi}_{AB} = \sum_{\ell}c_{\ell}\ket{\ell}_A \ket{H}_A \ket{-\ell}_B.
    \label{eq:HighDimOAM}
\end{equation}
where A and B are subscripts denoting each distinguishable photon, $c_{\ell}$ are complex scalar coefficients, $|H\rangle_A$ is the polarization of photon A (photon B's polarization has been omitted as it does not play a role in the final state) and $\ket{\ell}$ is the photonic state with OAM of $\ell\hbar$ per photon. The high-dimensional nature of the OAM-OAM correlations of the state is visualized through the experimentally measured OAM spiral bandwidth shown in Fig.~\ref{fig: MetaCharacter}\textbf{a}. Here the anti-diagonal matrix represents the expected anti-correlations between the two photons set by the OAM conservation of SPDC. One of the photons (in this case photon A) is sent through a J-plate metasurface, with a schematic of the metasurface design shown in Fig.~\ref{fig: MetaCharacter}\textbf{b}. The metasurface decouples orthogonal polarization states of incident photons and enables the mapping of independent phase profiles onto each polarization channel \cite{arbabi2015dielectric, li2023arbitrarily, mata2025tailoring}. For an input state $|P, \ell\rangle$ — where 
$|P\rangle = \cos\left(\frac{\theta}{2}\right)|H\rangle + e^{i\alpha}\sin\left(\frac{\theta}{2}\right)|V\rangle$ denotes an arbitrary polarisation state, the J-plate simultaneously imparts topological charges $m$ and $n$ to the horizontal (H) and vertical (V) polarization components, resulting in the transformation
$|P, \ell\rangle \xrightarrow{\text{metasurface}} \cos\left(\frac{\theta}{2}\right)|H,\ell+m\rangle + e^{i\alpha} \sin\left(\frac{\theta}{2}\right)|V,\ell+n\rangle$. 
In addition to the OAM phase, a linear phase gradient $K_G$ is superimposed onto both components, imparting transverse momentum for beam deflection. This separates the desired modulated field from unmodulated zero-order light. The total spatial phase profiles for the H and V channels are thus defined as
\begin{eqnarray}
    \Phi_H (x, y) &=& \Phi_m(x, y) + \Phi_G(x) \\
    \Phi_V (x, y) &=& \Phi_n(x, y) + \Phi_G(x),
\end{eqnarray}  
where $\Phi_m = -m\cdot\text{atan2}(y, x)$, $\Phi_n = -n\cdot\text{atan2}(y, x)$, and $\Phi_G = K_G\cdot x$. As an example, the phase profile superposition of OAM ($m=0 $ and $n=1$) with linear grating are illustrated in the lower panel of Fig.~\ref{fig: MetaCharacter} \textbf{b} to show the encoding of the two components. After passing photon A through the J-plate (photon A's polarization is first rotated to diagonal polarization) the resultant state transforms as  
\begin{eqnarray}
    \ket{\Psi}_{AB} &=& \sum_{\ell}c_{\ell}\ket{\ell}_A\ket{H}_A \ket{-\ell}_B \nonumber \\
    \xrightarrow{\text{J-plate}}  \ket{\Psi}_{AB} &=& \frac{1}{\sqrt{2}}\sum_{\ell}c_{\ell}\ket{\ell+m}_A\ket{H}_A \ket{-\ell}_B \nonumber \\ && + \frac{1}{\sqrt{2}}\sum_{\ell}c_{\ell}\ket{\ell+n}_A\ket{V}_A \ket{-\ell}_B \nonumber \\
    &=&\frac{1}{\sqrt{2}}\left(\ket{\Psi}_{AB,m}^H + \ket{\Psi}_{AB,n}^V\right).
\label{eq:HighDimPolSpiral}
\end{eqnarray}
where the initial high-dimensional, anti-correlated OAM-OAM state splits into two high-dimensional states, $\ket{\Psi}_{AB,m}^H,\ket{\Psi}_{AB,n}^V$ each labelled by an orthogonal polarization state of photon A, that have been shifted with respect to the initial state by $m$ and $n$, respectively. An example of such a transformation is shown in Fig.~\ref{fig: MetaCharacter}, passing a high-dimensional state through a metasurface with $(m,n) = (0,1)$. After the metasurface transformation, we see that while the spiral bandwidth of $\ket{\Psi}_{AB}^H$ has remained the same as $\ket{\Psi}_{AB}$, the spiral bandwidth of $\ket{\Psi}_{AB}^V$ has shifted by 1 element, as illustrated in Fig.~\ref{fig: MetaCharacter}\textbf{c}. The spiral bandwidth of the combined state then appears as two adjacent anti-diagonal lines.\\ 
If the state is labelled by the OAM of photon A (visualized as independent rows of the spiral bandwidth), the multi-dimensional state is revealed 
\begin{equation}
    \ket{\Psi}_{AB} = \frac{1}{\sqrt{2}}\sum_{\ell'} \ket{\ell'}_A \ket{\phi}^{\ell'}_{AB},
\label{eq:MultiDimState}
\end{equation}
where $\ket{\phi}^{\ell'}_{AB} = c_{-\ell'-m}\ket{H}_A\ket{-\ell'-m}_B + c_{-\ell'-n}\ket{V}_A\ket{-\ell'-n}_B$ are 2D hybrid states with $\ell' = \ell+m$ as desired. In this way, the multi-dimensional state takes the form of a superposition  of hybrid states labelled by the OAM of photon A, therefore existing in parallel.\\ 

\section*{Experimental setup and metasurface characterization}

\noindent The experimental setup used to generate, detect and switch between different non-local topologies is shown in Fig.~\ref{fig: MetaCharacter} \textbf{d}. A pair of OAM-OAM entangled photons of wavelength 810 nm were produced through an SPDC process by pumping a Type-I non-linear crystal (NC) with a 405 nm wavelength pump laser with a Gaussian profile. Any residual pump photons were filtered by an 810 nm wavelength bandpass filter (BPF). The entangled photons were then sent through separate paths in our setup for final transformation and measurement. Photon A was sent through a half-wave plate modifying its polarization state to diagonal, $\ket{D}$, followed by a polarization-dependent transformation enacted by the J-plate metasurface. The metasurface phase profiles were implemented using a circular-aperture metasurface (1.2 mm in diameter), where dielectric nanopillars imposed the desired polarization-dependent phase shifts through subwavelength form birefringence. The constituent meta-atoms are amorphous silicon cuboid nanopillars, which can be regarded as truncated rectangular waveguides. By tuning their transverse dimensions (width and length), the nanopillars exhibit strong geometric birefringence, enabling independent phase control spanning $0-2\pi$ for two orthogonal linear polarizations. At the operating wavelength of 810 nm, the nanopillars were designed with a fixed height of 500 nm and a lattice periodicity of 400 nm, ensuring subwavelength operation and suppressing higher-order diffraction. Using rigorous coupled-wave analysis, we constructed a 16×16 library of nanopillars that covered the full phase space for both H and V components. 
The metasurface was fabricated using standard electron-beam lithography followed by reactive-ion etching. 
 The high structural fidelity and the size-dependent geometry of the nanopillars are confirmed by tilted scanning electron microscopy (SEM) images (inset in Fig.~\ref{fig: MetaCharacter}\textbf{b}).\\
The spatial light modulator (SLM A) coupled with the single mode fibre (SMF A) was used to switch between different hybrid states within our larger multi-dimensional hybrid entangled state. The state was then measured by performing joint polarization and OAM projections on photon A and B, respectively. Photon A's polarization was measured using a set of polarization optics (POs) and photon B's OAM detected using a joint mode projection setup consisting of an SLM (SLM B) and SMF (SMF B) with the probability outcomes, $C_i^M$, summarized in a quantum state tomography (QST) dataset.\\

\begin{figure*}[t]
\centering\includegraphics[width=1\linewidth]{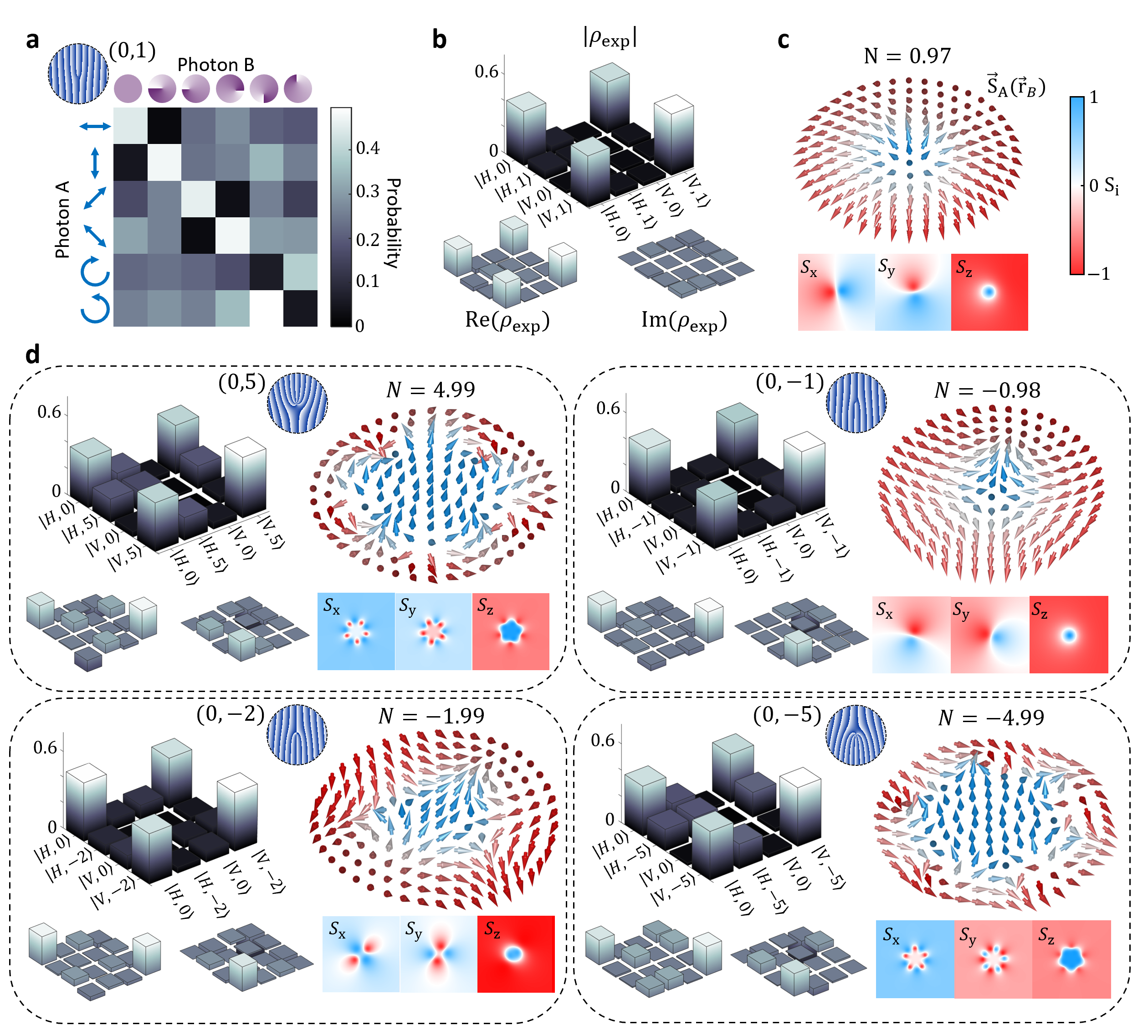}
\caption{\textbf{Various Metasurface assisted topologies.} \textbf{a}, Measured quantum state tomography and \textbf{b}, reconstructed absolute, real and imaginary components of the density matrix for the state, $|H\rangle_A|0\rangle_B + |V\rangle_A|1\rangle_B$, along with its \textbf{c}, quantum Stokes parameters, $\{S_x,S_y,S_z\}$, and quantum Stokes vector $\vec{S}(\vec{r}_B)_A$ created using the metasurface with $(m,n) = (0,1)$.  \textbf{d}, corresponding density matrix components, quantum Stokes components and Quantum Stokes vector field for states produced using metasurfaces with $(m,n) = (0,5),(0,-1),(0,-2) \;\text{and}\; (0,-5)$.} \label{fig: DiffMeta}
\end{figure*}

\noindent The J-plates used in this work were then characterized by analysing the 2D hybrid states produced with SLM A set to project onto the fundamental Gaussian mode. Additionally, a full Stokes polarimetric analysis was performed on the produced classical vector beams after probing them with a single Gaussian beam with these results shown in the accompanying Supplementary Information. In this scenario the measured 2D hybrid states are obtained by applying the projection operator $\ket{\ell=0}_A{}_A\bra{\ell=0}$ onto the state given in Eq.~\eqref{eq:MultiDimState} resulting in the state 
\begin{equation}
    \ket{\Psi}_{AB} = \frac{1}{\sqrt{2}}\left(\ket{H}_A \ket{m}_B + \ket{V}_A \ket{n}_B \right),
    \label{eq:2DimHybrid}
\end{equation}
Here, the pair of metasurface indices, $(m,n)$, uniquely select the resultant hybrid state of the system. As a first illustrative example we measured the state $\ket{\Psi}_{AB} = \frac{1}{\sqrt{2}}\left(\ket{H}_A \ket{0}_B + \ket{V}_A \ket{1}_B \right)$ generated using a J-plate with $(m,n) =(0,1)$. The measured QST for this state is shown in Fig.~\ref{fig: DiffMeta} \textbf{a} with the components of its associated reconstructed density matrix, $\rho$, depicted in Fig.~\ref{fig: DiffMeta} \textbf{b}. The density matrix for the measured state was reconstructed by minimizing $\chi^2=\sum_i^N{\frac{(C^M_i-C^P_i(\tilde{\rho}))^2}{C_i^P(\tilde{\rho})}}$ where $C_i^P(\tilde{\rho})$ are the probability outcomes for a predicted density matrix $\tilde{\rho}$. Further details of this procedure can be found in the supplemental information of \cite{ornelas2025topological}. The quality of the state is accessed through computation of the concurrence, $C$, fidelity, $F$, and purity, $\gamma$, (defined in the accompanying supplementary information) with values for each of the J-plate samples summarized in Table~\ref{table:EntWit_N}. The topology of the state was then calculated by first computing the quantum Stokes vector field, $\vec{S} = (S_x, S_y, S_z)$, where each component is given by $S_i = \text{Tr}\left( \sigma_i \rho(\vec{r})\right)$ and $\rho(\vec{r}) = \bra{\vec{r}}\rho\ket{\vec{r}}$ is the spatially-varying density matrix constructed using the relation $\langle\vec{r}|\ell\rangle = \text{LG}_{\ell}^{p=0}(\vec{r})$ where $\text{LG}_{\ell}^{p=0}(\vec{r})$ is the Laguerre gaussian function with azimuthal index $\ell$ and radial index $p=0$. A full representation of the quantum Stokes vector field for our measured state is shown in Fig.~\ref{fig: DiffMeta} \textbf{c}, depicting a Ne\'el-type texture. The topological number of the state is then quantified using
\begin{equation}
    N = \frac{1}{4\pi}\int_{R^2} \vec{S}\cdot\left(\partial_x\vec{S} \times \partial_y\vec{S}\right) \, d\vec{r}
    \label{eq:SkyNum1}
\end{equation}
which was calculated to be $N=0.97 \pm 1.92\times10^{-5}$ for our state, confirming the first quantum skyrmion generated using a metasurface. Altering the metasurface parameters $(m,n)$ can then directly control the skyrmionic topology. 
This can be seen if we compute the skyrmion number for the state given in Eq.~\eqref{eq:2DimHybrid} which results in 
\begin{equation}
    N = \text{sign}(|n|-|m|)(n-m) = pv,
    \label{eq: SkyNumMeta}
\end{equation}
where the integers $p = \text{sign}(|n|-|m|)$ and $v=n-m$ are known as the polarity and vorticity, respectively. Therefore, the J-plate can be used to alter the topology of the state to any topological class, as well as the texture of the topology, simply by varying the parameters $(m,n)$. In particular by altering the magnitude and sign of the vorticity of each state by designing metasurfaces with $(m,n) = (0,k)$ where $k\in\{-5,-2,-1,1,2,5\}$, we achieved states belonging to 6 distinct topological classes characterized by the measured topological numbers $N=\{-4.99, -1.99, -0.98, 0.97, 1.99, 4.99\}$ with the density matrices and quantum Stokes vector fields shown in Fig.~\ref{fig: DiffMeta}\textbf{d} for $n=\{5, -1, -2, -5\}$ and the entanglement witnesses summarized in Table \ref{table:EntWit_N}. The errors in all computed values from the density matrix are calculated by propagating statistical Poissonian errors in photon-count rates via Monte-Carlo simulation of the experiment. For our skyrmion states the variation in the sign and magnitude of the vorticity is seen as a change in the rotation direction of the texture about the $S_z$ axis ($+1$ if vectors rotate in the same direction as the changing azimuthal angle and $-1$ otherwise) and the change in the number of rotations performed by $\vec{S}_A(\vec{r}_B)$ about the $S_z$ axis, respectively. In each measured case, the polarity has been maintained ($+1$) with  $\vec{S}_A(\vec{r}_B)$ pointing up in the centre and down away from the origin.\\

\begin{table}[h!]
\centering
\caption{Summary of quantitative measurements for different J-plates characterized by integers $(m, n)$; Fidelity ($F$), Purity ($\gamma$), Concurrence ($C$)) and Skyrmion number ($N$).}
\begin{tabular}{|c|c|c|c|c|}
\hline
$(m, n)$ & C ($\pm 10^{-4}$) & F ($\pm 10^{-4}$) & $\gamma$ ($\pm 10^{-4}$) & $N$ ($\pm 10^{-4}$) \\
\hline
(0,1) & $0.76 \pm 4.30$ & $0.87 \pm 1.30$  & $0.78 \pm 3.20$ & $0.97\pm 0.25$\\
\hline
(0,2) & 0.82 $\pm 1.5$ & 0.80 $\pm 0.41$ & 0.81 $\pm 0.94$ & 1.99 $\pm 0.01$\\
\hline
(0,5) & 0.74 $\pm 0.65$  & 0.78 $\pm 2.05$ & 0.79 $\pm 3.40$ & 4.99 $\pm 0.19$\\
\hline
(0,-1) & 0.73 $\pm 2.82$  & 0.81 $\pm 0.57$ & 0.73 $\pm 0.91$ & -0.97 $\pm 0.19$\\
\hline
(0,-2) & 0.73  $\pm 3.19$ & 0.84 $\pm 0.86$ & 0.79 $\pm 1.70$ & -1.99 $\pm 0.02$\\
\hline
(0,-5) & 0.75 $\pm 6.90$  & 0.76 $\pm 2.20$ & 0.78 $\pm 4.60$ & -4.99$\pm 0.25$ \\
\hline
\end{tabular}
\label{table:EntWit_N}
\end{table}

\section*{Multi-level topological state generation}
\begin{figure*}[t]
\centering\includegraphics[width=1\linewidth]{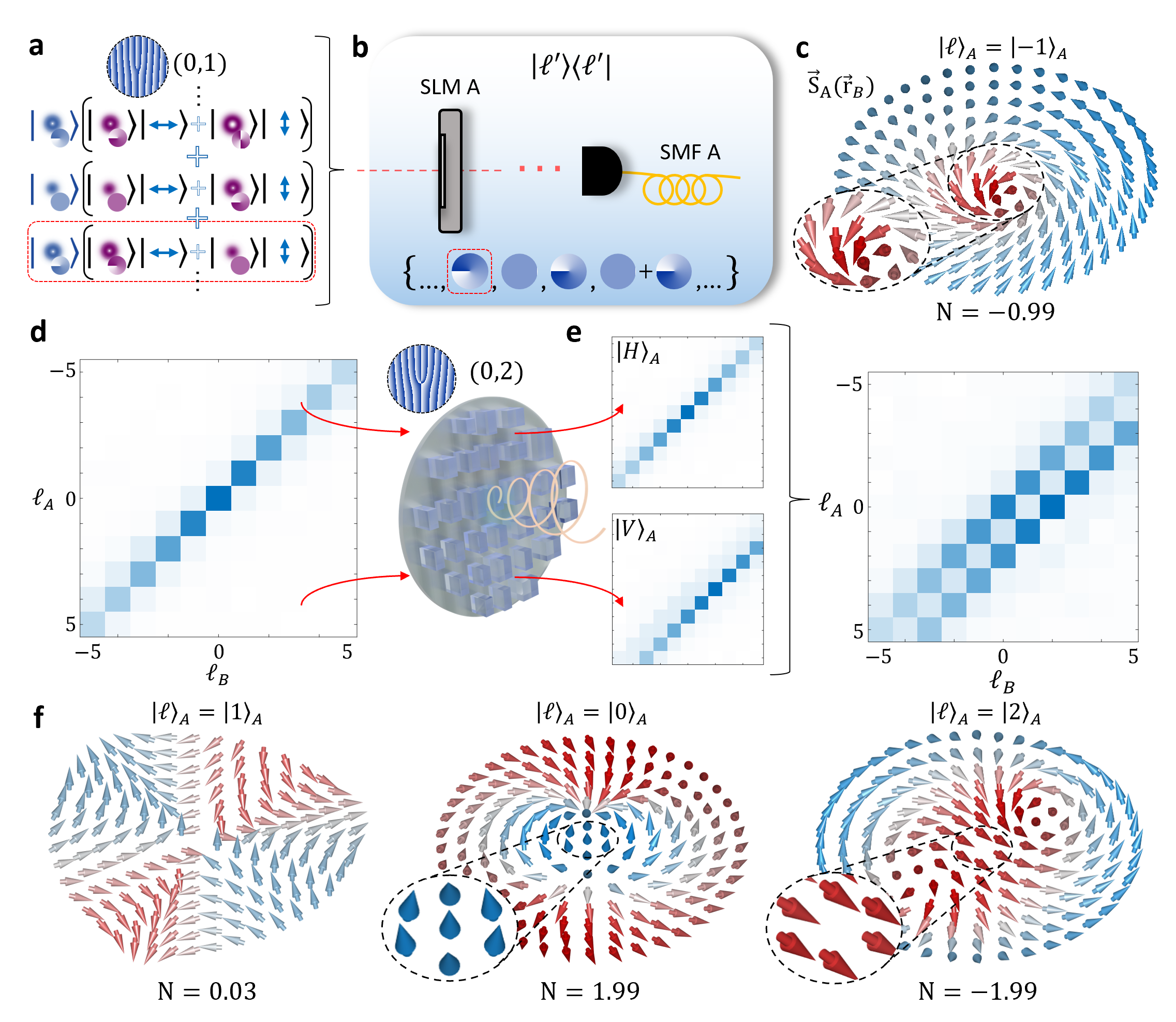}
	\caption{\textbf{Multi-Skyrmion state}. 
    \textbf{a} Example of multi-dimensional hybrid entangled state after passing high-dimensional OAM-OAM entangled state through J-plate metasurface with $(m,n) = (0,1)$. \textbf{b}, Projection of a particular hybrid state is done by displaying the conjugate OAM of photon A on SLM A and passing that photon through a SMF. For example, selection of the \textbf{c} OAM state $\ket{-1}_A$, collapses the system to a hybrid state possessing a flipped polarity and Skyrmion number when compared to the selection of $\ket{0}_A$. \textbf{d}, OAM spiral bandwidth of initial high-dimensional state. \textbf{e}, OAM spiral bandwidth of multi-dimensional state showing horizontal, $\ket{H}_A$, and vertical, $\ket{V}_A$, polarized components that combine to give the spiral bandwidth of the full spectrum. \textbf{f}, Three distinct quantum topologies measured from three embedded 2D subspaces within the multi-dimensional state.}
	\label{fig: HighDimEnhance}
\end{figure*}

\noindent We now consider projecting onto different hybrid states within our multi-dimensional state. These hybrid states take the form
\begin{equation}
    \ket{\phi}^{\ell'}_{AB} = c_{-\ell'-m}\ket{H}_A\ket{-\ell'-m}_B + c_{-\ell'-n}\ket{V}_A\ket{-\ell'-n}_B,
    \label{eq: HybridOAMProj}
\end{equation}
with normalized coeffecients, $|c_{-\ell'-m}|^2+|c_{-\ell'-n}|^2 = 1$. An illustrative example is shown in Fig.~\ref{fig: HighDimEnhance} \textbf{a}, where the J-plate chosen here is characterized by $(m,n) = (0,1)$. Here, the roles of the J-plate metasurface and high-dimensional entangled state are revealed. The J-plate metasurface sets the separation of the OAM modes over the entire multi-dimensional space whereas high-dimensional entanglement acts as a resource allowing for the construction of several 2D hybrid states in parallel. This was shown in Fig.~\ref{fig: MetaCharacter} \textbf{a,c} 
for the metasurface with $(m,n) = (0,1)$. There the resultant state could be seen exhibiting correlations between polarization and OAM within multiple distinct 2D hybrid states, with the separation of the horizontal and vertical spiral bandwidths set by the imparted OAM of the J-plate metasurface. The selection of different hybrid states then involves projecting onto different OAM modes on photon A. We achieve this by displaying the hologram of the conjugate OAM mode on SLM A which in conjunction with SMF A, performs the desired OAM mode projection on photon A, as depicted in Fig.~\ref{fig: HighDimEnhance} \textbf{b}. Evaluating the topological number for the state given in Eq.~\ref{eq: HybridOAMProj} yields
\begin{equation}
    N_{\ell'} = \text{sign}(|n-\ell'|-|m-\ell'|)(n-m) = p_{\ell'}v.
\end{equation}
Therefore, the chosen OAM channel $\ell'$ modifies the polarity of the output state. Under this paradigm, if $v$ is odd, then there are 2 possibilities for the topological number, $N_{\ell'} \in \{-v,v\}$. An example of this polarity flip is depicted by the quantum Stokes vector, $\vec{S}_A(\vec{r}_B)$, in Fig.~\ref{fig: HighDimEnhance} \textbf{c}, for the metasurface with $(m,n) = (0,1)$ after projecting photon A onto the OAM mode $\ket{-1}_A$, with $N = -0.99 \pm 0.17\times10^{-4}$. Comparing this result to that shown in Fig.~\ref{fig: DiffMeta} \textbf{c}, a clear flip in the polarity is visible, exhibited by the flip in orientation of the vector field at the origin and periphery. Furthermore, if $v$ is even then there are 3 possibilities for the topological number $N_{\ell'} = \{-v,0,v\}$. Here the additional topological classification, $N=0$, is made possible by the existence of an anti-symmetric state, $\ell_B\in\{-\ell,\ell\}$, within the multi-dimensional state, which has a polarity of zero thereby setting the Skyrmion number to zero. An example of this is shown in Fig.~\ref{fig: HighDimEnhance} \textbf{d-f} for the metasurface with $(m,n) = (0,2)$. Here the initial OAM-OAM correlations, shown in Fig.~\ref{fig: HighDimEnhance} \textbf{d}, are transformed by the metasurface into the two high-dimensional states, $\ket{\Psi}_{AB,0}^H$ and  $\ket{\Psi}_{AB,2}^V$ with shifted OAM correlations as depicted in Fig.~\ref{fig: HighDimEnhance} \textbf{e}. When projecting onto the three distinct OAM states for photon A, $\ket{\ell}_A\in\{\ket{1}_A,\ket{0}_A,\ket{2}_A\}$, three distinct topological numbers are revealed for a single state. These three experimentally measured quantum topologies are depicted in Fig.~\ref{fig: HighDimEnhance} \textbf{f}. 
The measured topological numbers for the three states are $N_{-1} = 0.03 \pm 0.43\times10^{-4}$, $N_{0} = 1.99 \pm 0.12\times10^{-5}$ and $N_{1} = -1.99 \pm 0.13\times10^{-4}$. Here, the topology derived from the state projected onto $\ket{\ell}_A=\ket{1}_A$ exhibits the same vorticity as the other states, except that the polarity is zero, as seen by the fact that the vectors remain pointing parallel to the ground plane across the spatial domain.\\

\noindent We have shown that while the metasurface was used to engineer OAM-polarization correlations from which our topology could be derived, the high-dimensionality of the state served to enhance this transformation so as to embed multiple topologies within a given multi-dimensional state. Next, we consider pushing this paradigm even further by evaluating the embedded topologies within higher-dimensional OAM Hilbert spaces by projecting onto multiple hybrid states at once. \\

\section*{Diverse topological state generation from superimposed hybrid states}

\begin{figure*}[t]
\centering\includegraphics[width=1\linewidth]{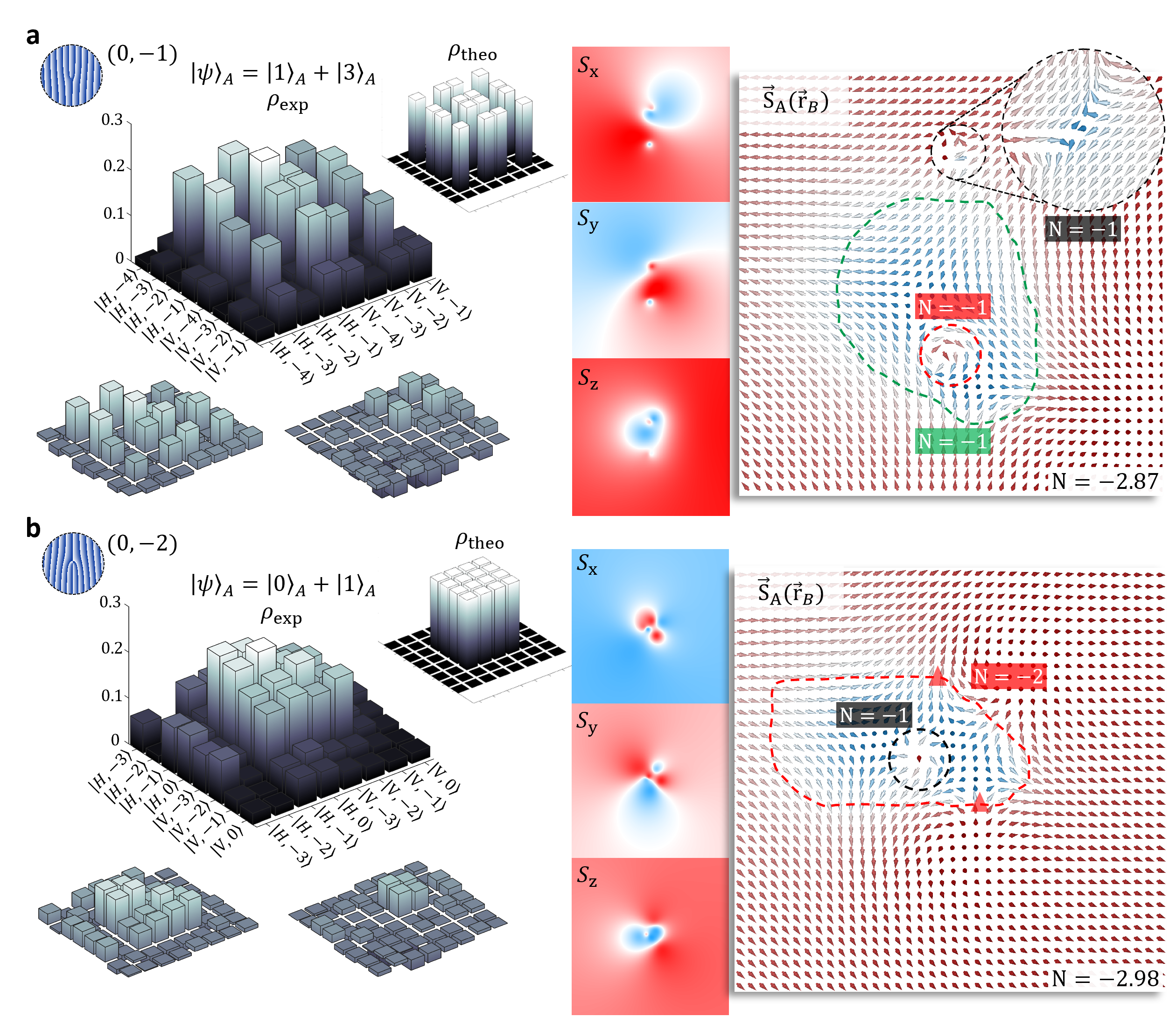}
	\caption{\textbf{Complex hybrid states possessing topological structures.} Measured density matrices, quantum Stokes components and Stokes vector fields for hybrid states generated by projecting onto superposition OAM states, \textbf{a} $|\ell\rangle_A = |1\rangle_A + |3\rangle_A $ and \textbf{b}, $|\ell\rangle_A = |0\rangle_A + |1\rangle_A$ for states generated by metasurfaces with $(m,n) = (0,-1)$ and $(m,n) = (0,-2)$, respectively. Topological features are highlighted by different coloured regions with highlighted topological numbers representing empirical numbers associated with each "local" topological feature. Calculated topological number of the full field is given in the inset at the bottom right of each vector field.} 
	\label{fig: HighDimMultiHybridState}
\end{figure*}

\noindent Up until this point, the hybrid states were selected by projecting onto single OAM states. However, our scheme is not limited to the OAM basis, in principle any spatial basis can be chosen thereby unlocking the true potential of our scheme with the possibility of generating more exotic topologies that coexist within the larger multi-dimensional state. This is achieved by displaying superpositions of OAM modes (chosen as an example of an arbitrary basis) on SLM A, thereby collapsing the system onto the state
\begin{eqnarray}
    \ket{\Omega}_{AB} = \sum_{\ell'}  \ket{\phi}^{\ell'}_{AB} = &&\sum_{\ell'}(c_{-\ell'-m}\ket{H}_A\ket{-\ell'-m}_B \nonumber \\
    &&+ c_{-\ell'-n}\ket{V}_A\ket{-\ell'-n}_B ),
\end{eqnarray}
where $ \sum_{\ell'}\left(|c_{-\ell'-m}|^2 + |c_{-\ell'-n}|^2 \right) = 1$. A simple rearrangement of the above expression yields the following 
\begin{eqnarray}
    \ket{\Omega}_{AB} = &&\ket{H}_A\sum_{\ell'}c_{-\ell'-m}\ket{-\ell'-m}_B \nonumber \\
    &&+ \ket{V}_A\sum_{\ell'}c_{-\ell'-n}\ket{-\ell'-n}_B,
\end{eqnarray}
where each polarization state is multiplied against a superposition of OAM states. Whilst the entangled state remains 2-dimensional, the Hilbert space of the OAM DoF that we measure has increased allowing for the construction of more complex states possessing exotic topological structures. To illustrate this we consider projecting onto two hybrid states simultaneously and evaluating the resultant state and topology. As illustrative examples we picked J-plate metasurfaces with $(m,n)=(0,-1)$ and $(m,n)=(0,-2)$ and studied the resultant topology when projecting onto the OAM states $|\psi\rangle_A = \ket{1}_A + \ket{3}_A$ and $|\psi\rangle_A = \ket{0}_A + \ket{1}_A$ for each metasurface, respectively. The corresponding hybrid states measured here are given by
\begin{eqnarray}
    \ket{\Omega}_{AB} = &&\ket{H}_A\left(c_{-3}\ket{-3}_B + c_{-1}\ket{-1}_B \right) \nonumber \\
    &&+ \ket{V}_A\left(c_{-4}\ket{-4}_B + c_{-2}\ket{-2}_B \right),
\end{eqnarray}
and 
\begin{eqnarray}
    \ket{\Omega}_{AB} = &&\ket{H}_A\left(c_{0}\ket{0}_B + c_{-1}\ket{-1}_B \right) \nonumber \\
    &&+ \ket{V}_A\left(c_{-2}\ket{-2}_B + c_{-3}\ket{-3}_B \right).
\end{eqnarray}
The measured density matrices, quantum Stokes parameters and Stokes vector fields are shown in Fig.~\ref{fig: HighDimMultiHybridState} \textbf{a} and \textbf{b} for each metasurface, respectively. Here the measured density matrices are $8\!\times\!8$ matrices corresponding to the increase in the dimensionality of the measured OAM Hilbert space. The measured fidelities and purities are $F=0.77 \pm 0.02$, $\gamma=0.74 \pm 0.05$ for $(m,n)=(0,-1)$ and  $F=0.73 \pm 0.02$, $\gamma=0.71 \pm 0.02$ for $(m,n)=(0,-2)$. The corresponding quantum Stokes parameters and Stokes vector field reveal the exotic topological structures of these states. For $(m,n) = (0,-1)$ the measured topological number is given as $N=-2.87\pm0.38$ resulting from the presence of three distinct topological features highlighted in the vector field. The zoomed-in inset shows the existence of a saddle-like feature with a $2\pi$ rotation performed by the vectors along the equator of the Poincar\'e sphere, consistent with an anti-skyrmionic structure. The larger domain, highlighted in "green", reveals a nested skyrmionic structure where a Bloch-type Skyrmion, shown within the red circular region, with positive vorticity ($v=1$) and negative polarity ($p=-1$) is embedded within a larger Skyrmionic structure with the same Skyrmion number but with opposite signs for the vorticity and polarity. Each of these features contribute $-1$ to the total Skyrmionic number thereby yielding a number close to the expected integer result of $-3$.   
Furthermore, for $(m,n) = (0,2)$ the measured topological number is given as $N=-2.98\pm0.25$. Here, the wrapping number can be deduced from the nested skyrmionic structure found in the region enclosed by the red boundary. The first is the "localized" Ne\'el-type distribution isolated within the larger Skyrmionic topological structure highlighted by a black circular border in Fig.~\ref{fig: HighDimMultiHybridState}\textbf{b}.
The second contributing feature is the outer vector field possessing a vorticity of $v=-2$ depicted by the $4\pi$ anti-clockwise rotation performed by the vectors around the equator of the Poincare sphere along the path illustrated by the red boundary of the Stokes vector field shown in Fig.~\ref{fig: HighDimMultiHybridState}\textbf{b}. These topological features contribute $N=-1$ and $N=-2$ to the total wrapping number thereby resulting in the measured integer $N\approx-3$. It should be noted here that these measured topological numbers are outside of what could be achieved with these metasurfaces through projections onto individual OAM states alone, thereby highlighting the existence of a larger diverse sets of topologies achievable with our high-dimensional enhancement scheme.  In future it should be possible to tailor these textures through entanglement engineering for topological entanglement witnesses \cite{nothlawala2026remote} and quantum versions of skyrmion bags \cite{schwab2026tunable}.

\section*{Discussion and Conclusion}
 
\noindent Quantum skyrmions offer a promising route towards robust information transmission, however, the techniques available for their generation remain in their infancy. Recent advances have enabled more compact and minitiarised implementations using metasurfaces, particularly in classical optical settings, while there is a slow progression towards their implementation in the quantum domain. However, typical approaches implemented within coherent beams and single photon optical platforms reach a natural bottleneck in topological information storage. 
Here, by interfacing high-dimensional entanglement with a J-plate metasurface, we establish a route to generating states that host multiple embedded topologies, by exploiting the often overlooked parallelism embedded within high-dimensional biphoton entangled states. 
Importantly, we represent these multi-dimensional states as distinct 2D hybrid states labelled by the OAM of one photon. This formulation renders our scheme naturally compatible with OAM mode sorters \cite{fickler2014interface, fontaine2019laguerre, liu2021sorting, ruffato2018compact, ruffato2019non}, enabling the decomposition of the multi-dimensional state into separate OAM-path channels and providing a practical avenue toward multi-channel topological networks. Furthermore, we demonstrate access to an expanded and more exotic topological landscape by moving beyond the OAM basis to arbitrary spatial mode bases. This extension aligns with recent advances in mode-sorting techniques for general spatial modes \cite{defienne2020arbitrary, kupianskyi2023high}, opening the door to richer classes of topological quantum states for higher-capacity information processing and communication networks.\\

\noindent In our analysis, we chose to investigate the Skyrmionic topology embedded within the non-local correlations between distant photons, however such topological features can also be found within the non-separable intrisic polarization and OAM DoFs of each photon, thereby making our scheme compatible with the on-demand generation of single-photon topologies \cite{ma2025nanophotonic, koni2025dual} through heralding by OAM measurement. Furthermore, in our experiment, we employed a separate nonlinear crystal and J-plate metasurface. However, we anticipate that this generation platform can be made even more compact through the use of lithium niobate non-local metasurfaces, which have recently demonstrated the direct generation of spatially entangled photon pairs with enhanced pair-production rates \cite{zhang2022spatially}. By unifying high-dimensional state generation and spin–orbit coupling within a single device, such a platform could enable the realization of multi-dimensional states directly at the source, paving the way for ultra-compact and highly versatile quantum topology generation.\\     

In conclusion, we have introduced a scheme that improves upon traditional Skyrmion generation techniques by exploiting the parallelism offered by high-dimensional biphoton entanglement. This is achieved through interfacing high-dimensional OAM–OAM entanglement with a metasurface for the generation of multi-dimensional OAM-OAM-spin entangled states. Within this approach, topology emerges as a joint property of spin–OAM entangled photons, engineered through transmission of the one photon through a J-plate metasurface. Here, distinct topological states are coupled to the OAM of the single photon, giving rise to multiple coexisting quantum topologies embedded within a single quantum state that are only revealed upon measurement. 
While controlling the desired output state by spatial mode projections on one of the entangled photons, we have demonstrated the production of multiple topological states from a single metasurface device, having measured up to three distinct topological states by projecting onto OAM states. We further extended the scheme by projecting onto superpositions of OAM modes, enabling the generation of increasingly diverse and complex topological structures revealing a richer, previously hidden topological landscape within the multi-dimensional state space. Our results reveal a new capability when structured entangled photons are interfaced with structured matter, allowing for the direct enhancement of traditional topological generation schemes that make use of single photons or classical coherent beams as inputs. 
This paves the way for their integration into miniaturized platforms for robust, high-dimensional quantum communication and information processing.


\section*{Acknowledgements} 
P.O. and F.N. acknowledge funding from the Council for Scientific and Industrial Research under the HCD-IBS scholarship scheme. A.F and I.N acknowledge funding from the South African Quantum Technology Initiative (SA QuTI). R.B. acknowledges support from Elevate Scholarship, delivered by the Academy of Technological Sciences and Engineering (ATSE) and funded by the Australian Government Department of Industry, Science and Resources (DISR) and Department of Defence. H.R. acknowledges the Australian Research Council grant (grant nos. FT250100565); H. Y., and S.A.M. acknowledge the Australian Research Council grant (grant nos. DP250102064); S. A. M. acknowledges the Lee Lucas Chair in Physics; This work was performed in part at the Melbourne Centre for Nanofabrication (MCN) in the Victorian Node of the Australian National Fabrication Facility (ANFF). \\

\section*{Data availability}
Data supporting this study are available from the authors upon reasonable request. \\

\section*{Conflict of interest}
The authors declare no conflict of interest. \\

\section*{References}

\clearpage
\appendix

\setcounter{section}{0}
\setcounter{figure}{0}
\setcounter{table}{0}
\setcounter{equation}{0}
\setcounter{footnote}{0}
\renewcommand{\thesection}{S\arabic{section}}
\renewcommand{\thefigure}{S\arabic{figure}}
\renewcommand{\thetable}{S\arabic{table}}
\renewcommand{\theequation}{S\arabic{equation}}

\begin{widetext}

\section*{Supplementary: Classical Metasurface characterization}

To characterise the six metasurfaces used in this work, Stokes polarimetry was performed on the classical vector beams produced by sending a diagonally-polarized gaussian beam through each metasurface. This was then followed by an analysis of the topological of the beam to ensure that the metasurfaces would behave as intended. \\
 
 \noindent To obtain these measurements, a diode laser operating at a target wavelength of 810nm was expanded from a single-mode optical fibre and imaged onto each metasurface J-plate. As the metasurface modulates the phase of the vertically polarised component of the beam, the input polarisation was set to anti-diagonal using a linear polariser placed before the metasurface. Since the metasurfaces were designed with a grating and would perform a phase-only modulation on the beam, the vector beam was evaluated in the far-field with intensity measurements recorded for six polarisation projections allowing the full classical Stokes parameters to be reconstructed. The projected intensity measurements for each metasurface is shown in Fig.~\ref{fig:supp_intensity_measurements}, with the inset images showing the corresponding simulated projections. As expected the horizontal (H) and vertical (V) projections for each metasurface exhibit gaussian and doughnut-like amplitude profiles as expected. Furthermore, the projections corresponding to diagonal (D), anti-diagonal (A), right- (R) and left (L) circular projections exhibit points of null-intensity due to the interference of a Gaussian mode and an OAM mode with the number of such points correctly corresponding to the OAM imparted on the V polarization component.\\

\begin{figure*}[h]
\centering\includegraphics[width=0.8\linewidth]{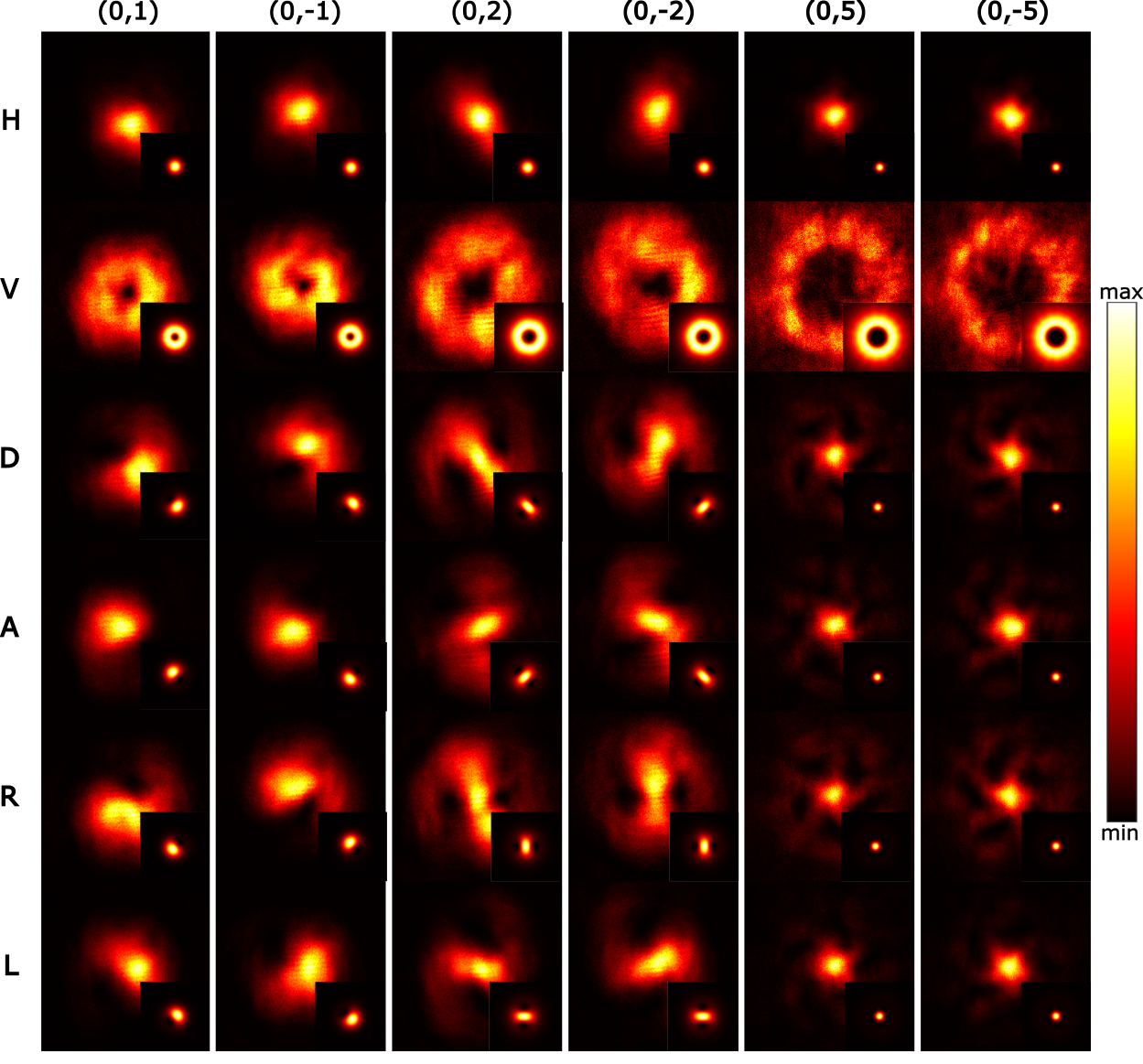}
	\caption{Classical metasurface polarisation projection intensity measurements and simulated metasurface responses under each projection (inset). Each row corresponds to one of the six metasurface J-plates investigated in this study. Far-field intensity distributions are shown for each polarisation projection. The corresponding OAM imparted by each metasurface is listed per row as pairs, corresponding to the OAM states imparted to the horizontal and vertical polarisation components, respectively.}
    \label{fig:supp_intensity_measurements}
\end{figure*}

\noindent From the measured far-field intensity projections the spatially-varying polarization structure can be discerned by computing the spatially varying Stokes parameters, $(S_0,S_1,S_2,S_3)$, which  
can be directly calculated from the measured intensity projections as
\begin{equation}
    \begin{bmatrix}
S_0\\
S_1\\
S_2\\
S_3
\end{bmatrix}
=
\begin{bmatrix}
H + V\\
H - V\\
D - A\\
R - L
\end{bmatrix}.
\label{eq:supp_stokescalc}
\end{equation}


The calculated Stokes parameters normalized against the maximum of $S_0$ are shown in Fig.~\ref{fig:supp_ClassicalStokes}, with the inset images showing the corresponding simulated metasurface responses. Comparison of the simulated and experimental Stokes fields shows good agreement in the overall structure. In particular, $S_1$, reveals a characteristic variation from $1$ to $-1$ with changing radius consistent with the cylindrical symmetry of the underlying fields. Furthermore, the number of lobes associated with the azimuthal polarisation phase variation of the beam is consistent between the simulated and experimental results indicating a strong agreement between the global topological structures of the measured and encoded beams.\\

\begin{figure*}[h]
\centering\includegraphics[width=0.8\linewidth]{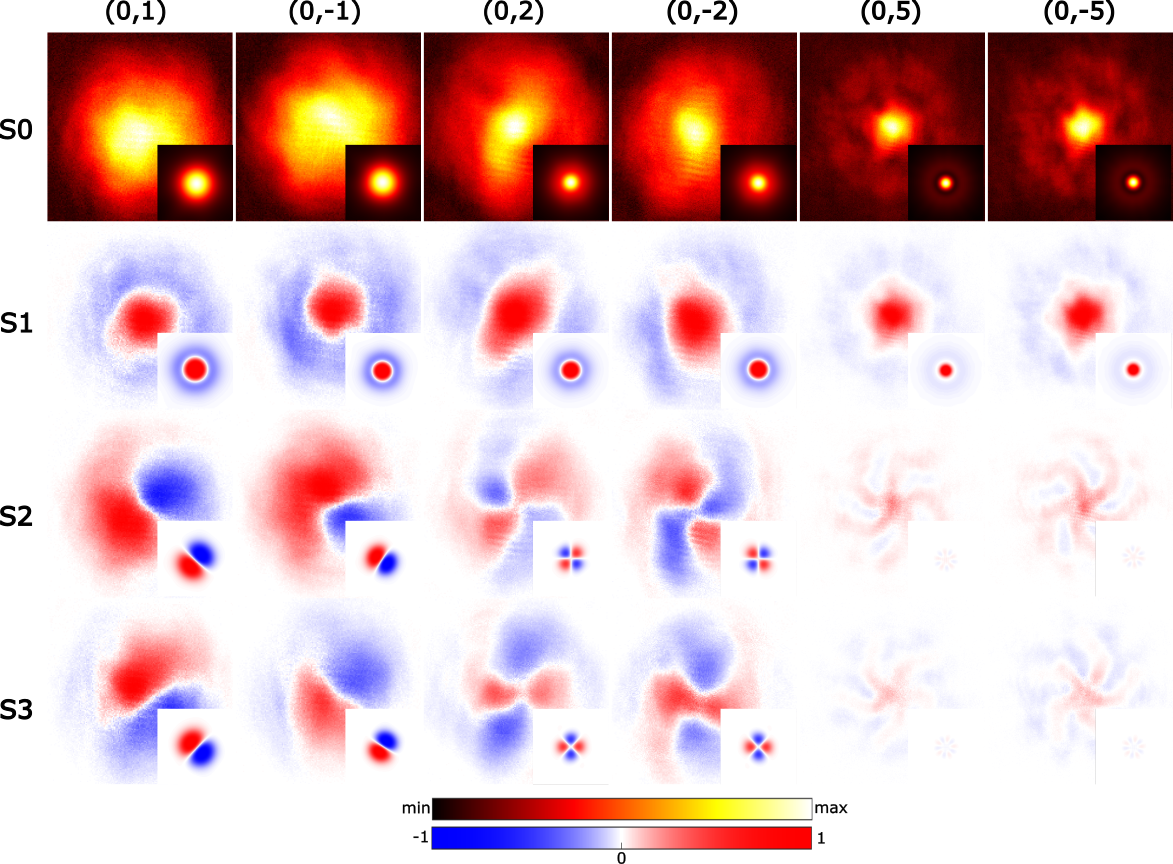}
	\caption{Classical experimentally derived  and simulated (inset) Stokes parameters for each metasurface. Each row corresponds to one of the six metasurface J-plates investigated in this study. The Stokes parameters for each point are calculated from the intensity projections with \ref{eq:supp_stokescalc}, and the $S_1,S_2,S_3$ shown are all globally normalised by dividing them by the maximum value of $S_0$ for each metasurface. The corresponding OAM imparted by each metasurface is listed per row as pairs, corresponding to the OAM states imparted to the horizontal and vertical polarisation components, respectively.}
    \label{fig:supp_ClassicalStokes}
\end{figure*}

The full state of polarization (SOP) for each beam is shown in Fig.~ \ref{fig:supp_beampolarisation}. The colormap varies with changes in $S_3$, thereby highlighting a full transition about the $S_1$ axis of the Poincar\'e sphere with each full $1$ (blue) to $-1$ (red) transition. Within each of these full transition regimes, highlighted by the partitions created using blue lines, the field fully covers the Poincar\'e sphere indicating a full wrapping of the Poincar\'e sphere. The number of observed transitions are consistent with the magnitude of the calculated Skyrmion number, $N$ (calculation method shown in the proceeding section), since we expect to observe $|N|$ full Poincar\'e sphere wrappings.

\begin{figure*}[h]
\centering\includegraphics[width=0.8\linewidth]{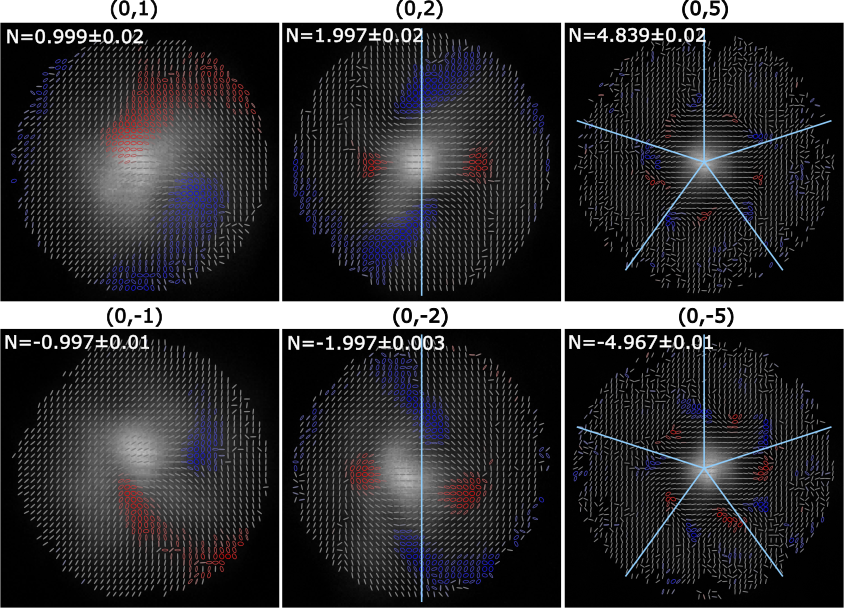}
	\caption{Shown are the total intensity responses from each metasurface J-plate (The $S_0$ field across the beam) labelled by the OAM imparted on the horizontal and vertical polarizations by the metasurface respectively, overlaid with ellipses indicating the local polarisation state across the beam. Circular polarisation states are coloured red or blue to denote right- and left-handedness, respectively. Radial lines are drawn at equal azimuthal spacing to partition the beam into a number of sectors equalling the number of times the Skyrmion field wraps the Poincare sphere.}
    \label{fig:supp_beampolarisation}
\end{figure*}

\section*{Supplementary: Extracting topological features from metasurface measurements}

Next the topology is evaluated for each metasurface. Here, care is taken to ensure an accurate evaluation of the Skyrmion number is carried out in the presence of noise, with more details provided in ref. \cite{peters2026extracting}.
As a point of comparison, the conventional integral definition of the Skyrmion number is used, defined as
\begin{equation}
    N=\frac{1}{4\pi}
\iint_{\mathbb{R}^{2}}
\mathbf{S}\cdot
\left(
\partial_x\mathbf{S}
\times
\partial_y\mathbf{S}
\right)
\,dx\,dy,
\label{eq:skynum}
\end{equation}

where $\mathbf{S} = [S_1(x,y), S_2(x,y), S_3(x,y)]$ is the normalised Stokes vector field across the beam. In practice, the integration is performed over a finite region, and as a result the experimentally calculated Skyrmion numbers are typically slightly lower in magnitude than the theoretical values, even under idealised conditions. Furthermore, this calculation depends strongly on the spatial gradients of the Stokes parameters across the beam, which become artificially large in regions of low signal-to-noise ratio, such as the low-intensity regions towards the edges of the beam. 
An approach to reduce the influence of noise on the Skyrmion number calculation is to restrict the integration to regions of the beam with intensity above a chosen threshold. The effect of thresholding on the calculated Skyrmion number using Eq.~\ref{eq:skynum} for the classical measurements of each metasurface is shown by the blue curves in Figure \ref{fig:supp_topologyanalysis}. When integrating across the full beam with zero threshold, corresponding to inclusion of all measured pixels, noisy regions contribute significantly and the resulting Skyrmion numbers are non-integer and typically deviate strongly from the designed values generated by the metasurfaces. As the intensity threshold fraction is increased, the calculated Skyrmion number briefly stabilises at values slightly lower in magnitude than the expected integers, as low signal-to-noise regions are filtered out. At higher threshold values, progressively higher-intensity regions are also excluded, causing the Skyrmion number to decay towards zero as less of the beam contributes to the integral. Overall, this approach provides a qualitative estimate of the Skyrmion number for each generated state. However, the presence of noise in the experimental data limits the ability to precisely quantify the Skyrmion states using this method.\\

\noindent In order to more accurately estimate the Skyrmion number for the generated states, we apply an alternative formulation that is mathematically equivalent but does not rely on the noisy spatial gradients introduced by the measurement process. This approach characterises the Skyrmion field using a contour integral, identifying polarisation singularities as signatures of the nontrivial topology of the beam \cite{mcwilliam2023topological, peters2026extracting}. The Skyrmion field components can be expressed as
\begin{equation}
    \Sigma_i = \frac{1}{2}\epsilon_{ijk}[\mathbf{S} \cdot (\partial_j \mathbf{S} \times \partial_k \mathbf{S})],
\end{equation}
where the Skyrmion number $N$ is given by the full integral

\begin{equation}
        N=\frac{1}{4\pi}
\iint_{\mathbb{R}^{2}}
\Sigma \cdot
d\mathbf{A}.
\end{equation}
This expression represents the general case of Equation \ref{eq:skynum}, where the previous form corresponds to the surface integral of the $z$-component of the Skyrmion field. The Skyrmion field $\Sigma$ is divergenceless, satisfying $\nabla \cdot \Sigma = 0$, and can therefore be written as the curl of another vector field, $\Sigma = \nabla \times \mathbf{v}$. Using this representation, the Skyrmion number can be equivalently expressed as the contour integral
\begin{equation}
    N = \frac{1}{4\pi}\oint_C \mathbf{v} \cdot dl,
    \label{eq:supp_contourmethod}
\end{equation}
where $C$ is a suitable contour in $\mathbb{R}^2$ that excludes the singularities of $\mathbf{v}$. This restriction on the contour allows us to specify a set of appropriate choices for the definition of $\mathbf{v}$ used in our analysis. To illustrate a suitable choice of $\mathbf{v}$ for our dataset, we first present an example of a form that would not be appropriate. If we define
\begin{equation}
    \mathbf{v}=-S_1\nabla\Phi, \; \text{where } \Phi=\text{arg}(S_2+iS_3),
\end{equation}
then we are identifying singularities in $S_1$. In the ideal case, these singularities correspond to points where $|S_1| = 1$, and hence exhibit a nontrivial winding in the local polarisation phase $\Phi$ around them. For all Skyrmion states generated using the metasurface J-plate, it can be seen from Fig.~\ref{fig:supp_ClassicalStokes} that the singularities in $S_1$ occur at the centre of the beam profile and as $r \rightarrow \infty$. In this case, to capture the Skyrmion number accurately, the contour should enclose the entire region up to $r \rightarrow \infty$, which in practice corresponds to integrating around the outer edge of the full image. However, the image boundary corresponds to a region of extremely low signal-to-noise ratio, which would lead to a highly noisy estimate of the Skyrmion number. Instead, a more appropriate choice of $\mathbf{v}$ is one that confines the singularities within a region of relatively high intensity ($S_0$), such as
\begin{equation}
    \mathbf{v}=-S_2\nabla\Phi, \; \text{where } \Phi=\text{arg}(S_3+iS_1),
\end{equation}
or
\begin{equation}
    \mathbf{v}=-S_3\nabla\Phi, \; \text{where } \Phi=\text{arg}(S_2+iS_3).
\end{equation}
While these two definitions are convenient for computation, it can be noted that rotations of the Poincaré sphere leave the Skyrmion number invariant. As a result, there exists a range of choices of $\mathbf{v}$ that preserve the singularities within the desired high signal-to-noise region. This admissible range was estimated empirically for each metasurface by identifying rotations that keep the singularities within regions exceeding a chosen intensity threshold of $S_0$. The resulting set of allowable orientations is shown as a band on the Poincaré sphere inset in each subfigure of Figure \ref{fig:supp_topologyanalysis}. This band corresponds to the region in which the plane defining $\Phi$ may be consistently chosen. For each threshold fraction of $\max(S_0)$, Skyrmion numbers were calculated using the contour integral method equivalent to Equation \ref{eq:supp_contourmethod} across 400 equally spaced choices of $\mathbf{v}$, corresponding to rotated $\Phi$ planes constrained within the identified bands. The resulting values are averaged and shown as the purple curves in Figure \ref{fig:supp_topologyanalysis}. The shaded region around each purple curve indicates the $\pm \sigma$ interval, where $\sigma$ is the standard deviation of the Skyrmion number across the 400 sampled Poincaré sphere rotations. For each metasurface, these plots reveal a threshold interval in which the standard deviation of the Skyrmion number obtained from the averaged contour method is approximately zero. This region of stability coincides with the stable region observed in the direct surface integral method, supporting the interpretation that this threshold interval corresponds to a regime in which the singularities are sufficiently well resolved to yield a robust measurement of the topological properties of the generated Skyrmion states. A horizontal line is fitted across these stable regions identified using the contour method, yielding Skyrmion numbers of ${0.999 \pm 0.02, -0.997 \pm 0.01, 1.997 \pm 0.003, -1.997 \pm 0.02, 4.839 \pm 0.02, -4.967 \pm 0.01}$ for the generated states.

\begin{figure*}[h]
    \centering

    \begin{minipage}{0.31\textwidth}
        \centering
        \includegraphics[width=\linewidth]{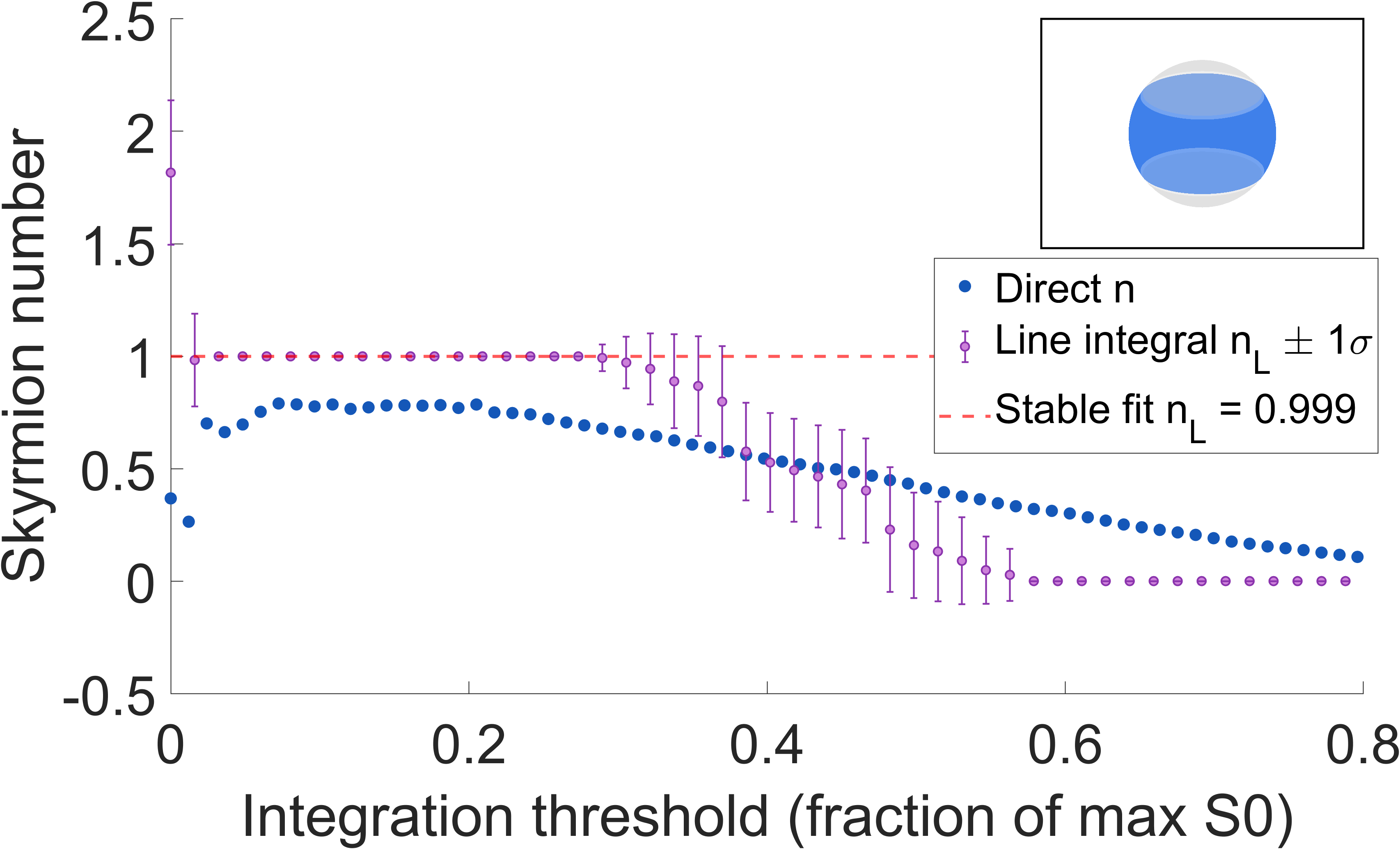}
    \end{minipage}
    \hspace{0.01\textwidth}
    \begin{minipage}{0.32\textwidth}
        \centering
        \includegraphics[width=\linewidth]{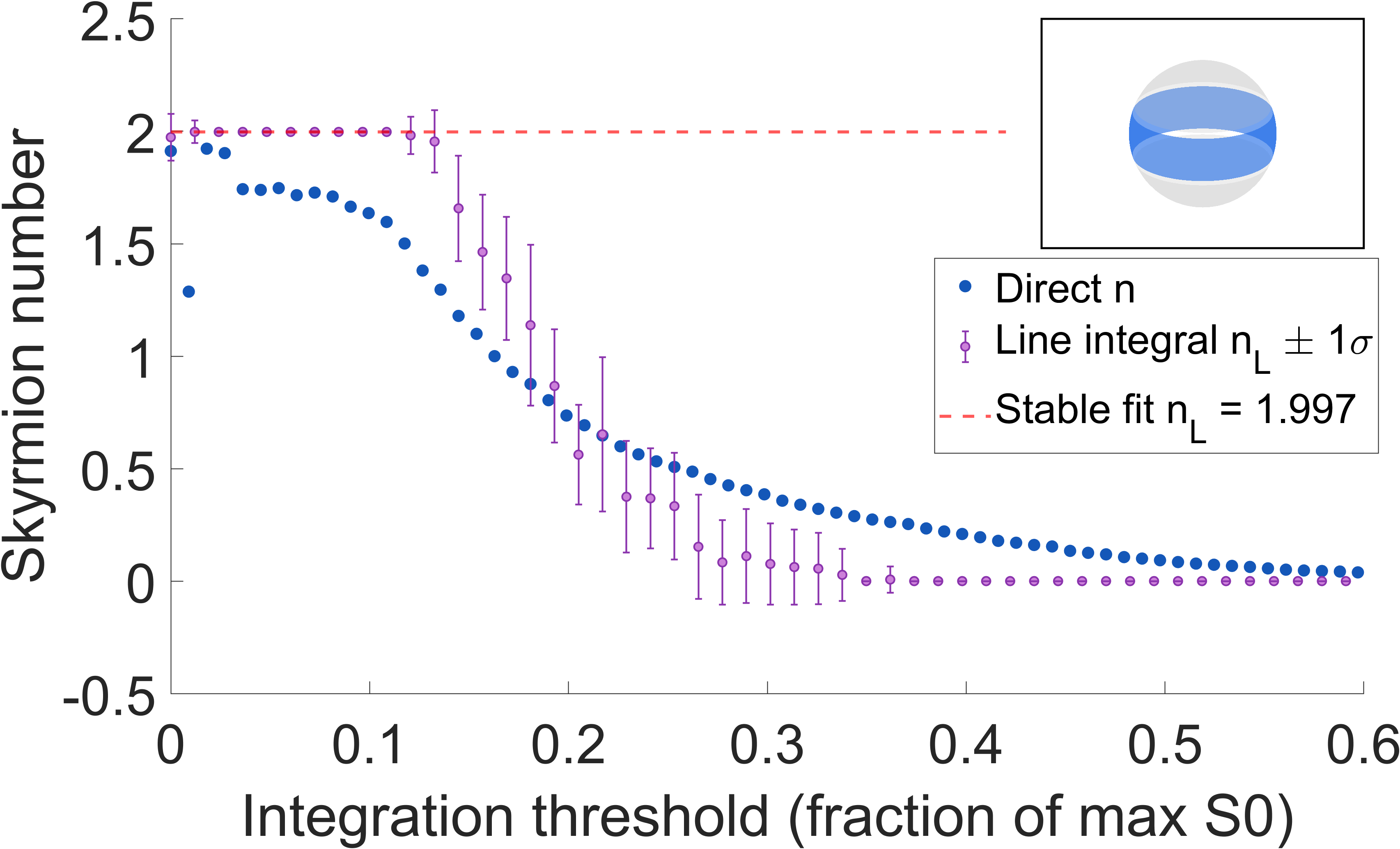}
    \end{minipage}
    \hspace{0.01\textwidth}
    \begin{minipage}{0.31\textwidth}
        \centering
        \includegraphics[width=\linewidth]{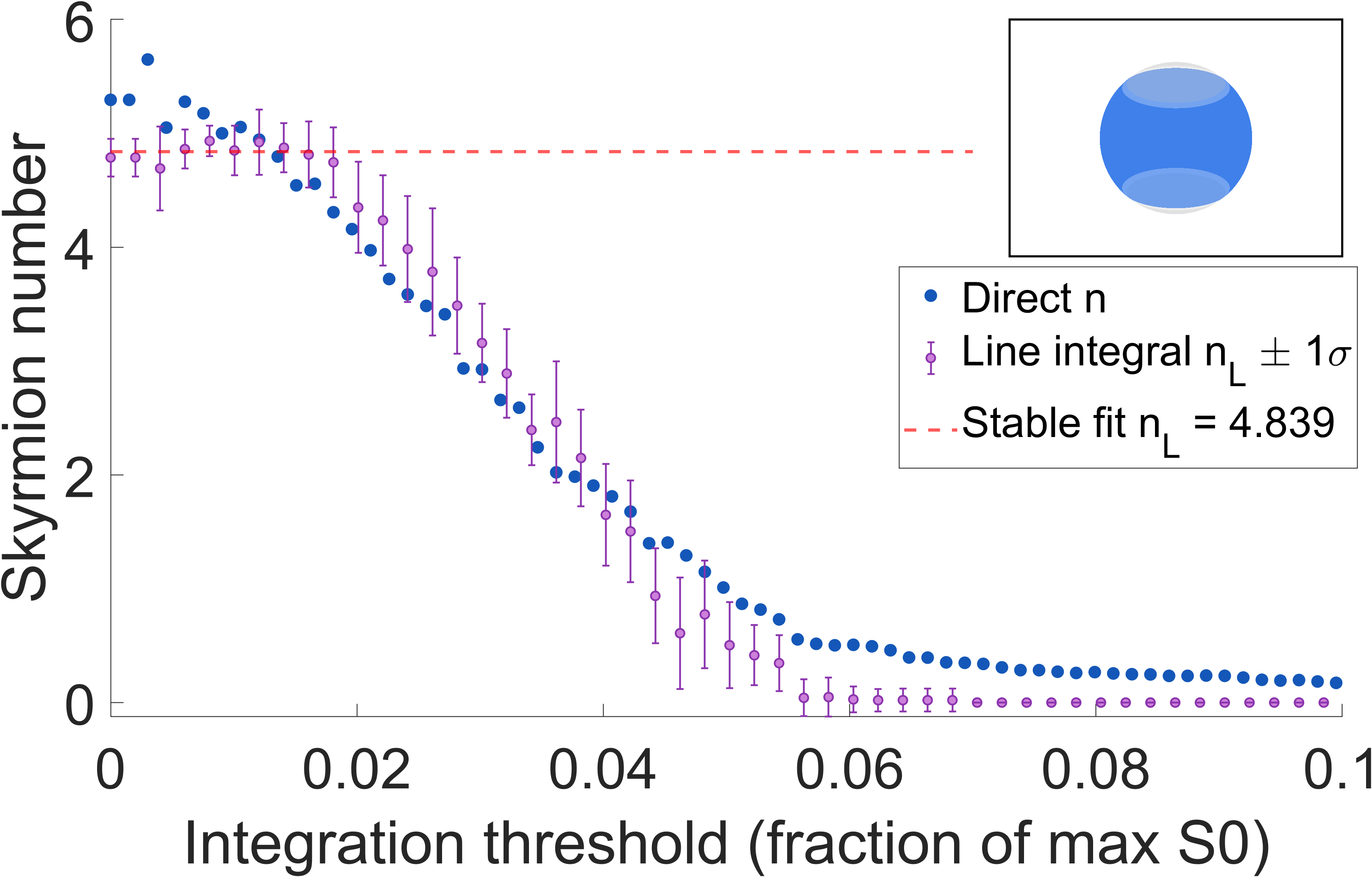}
    \end{minipage}

    \vspace{0.3em}

    \begin{minipage}{0.31\textwidth}
        \centering
        \includegraphics[width=\linewidth]{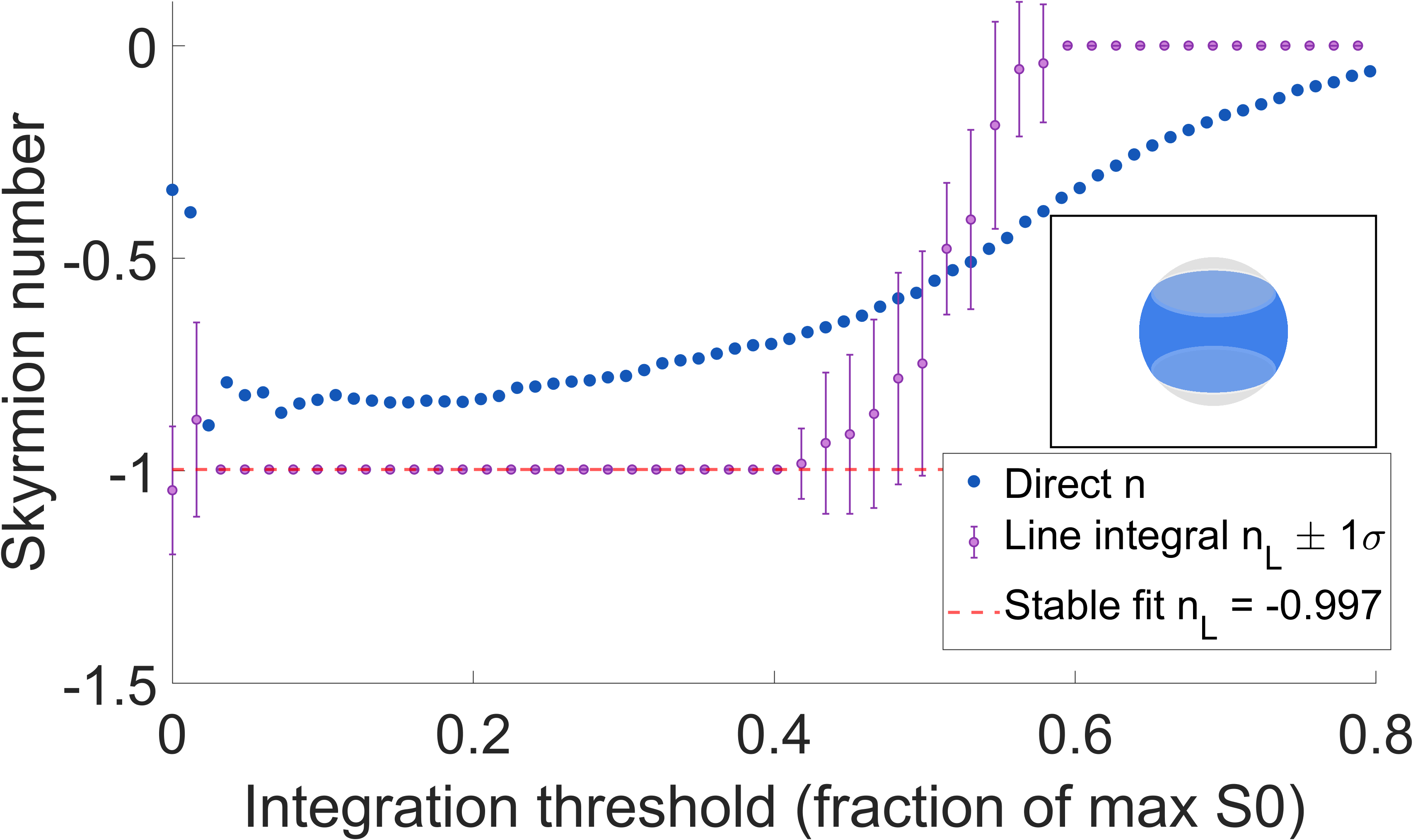}
    \end{minipage}
    \hspace{0.01\textwidth}
    \begin{minipage}{0.31\textwidth}
        \centering
        \includegraphics[width=\linewidth]{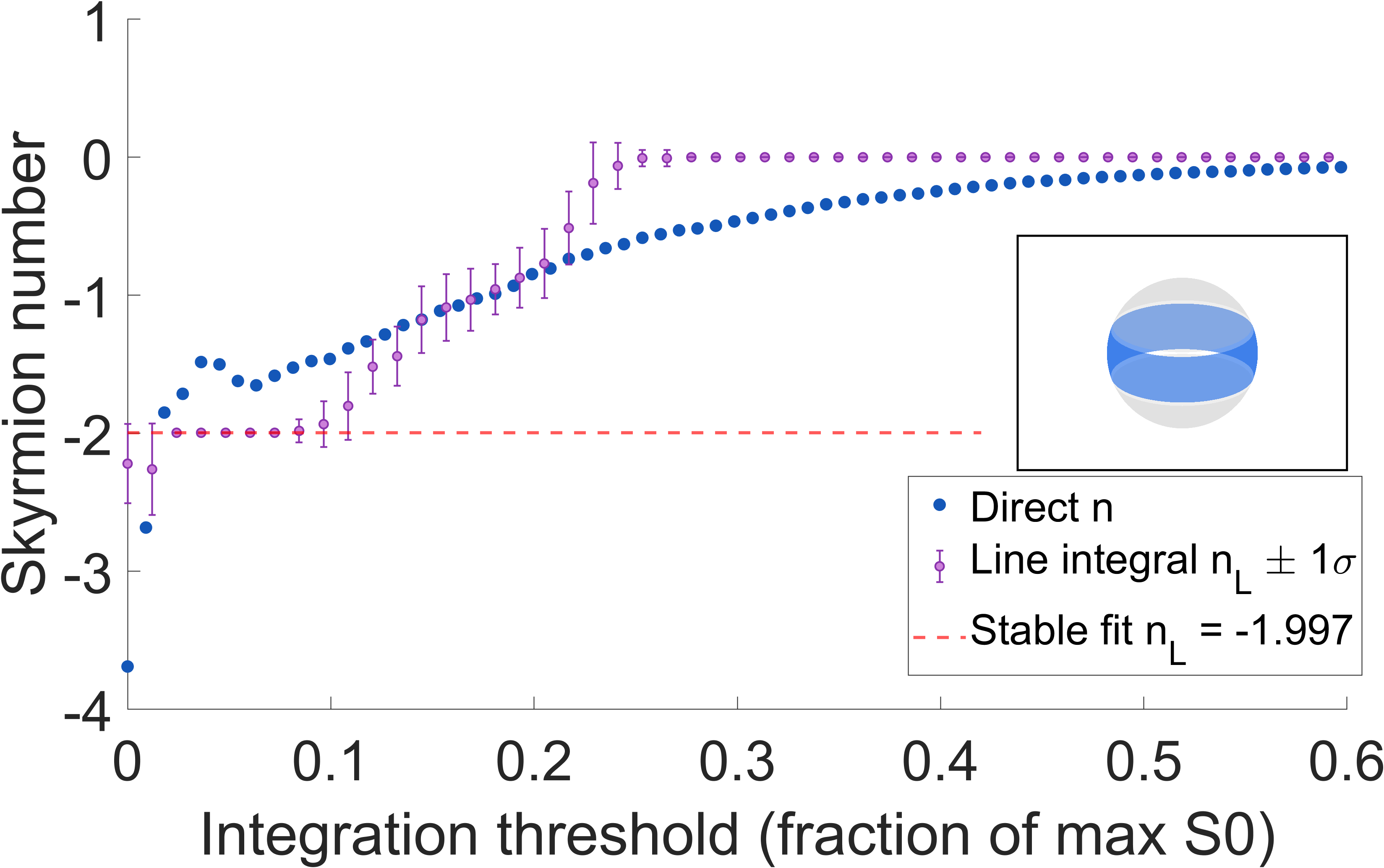}
    \end{minipage}
    \hspace{0.01\textwidth}
    \begin{minipage}{0.31\textwidth}
        \centering
        \includegraphics[width=\linewidth]{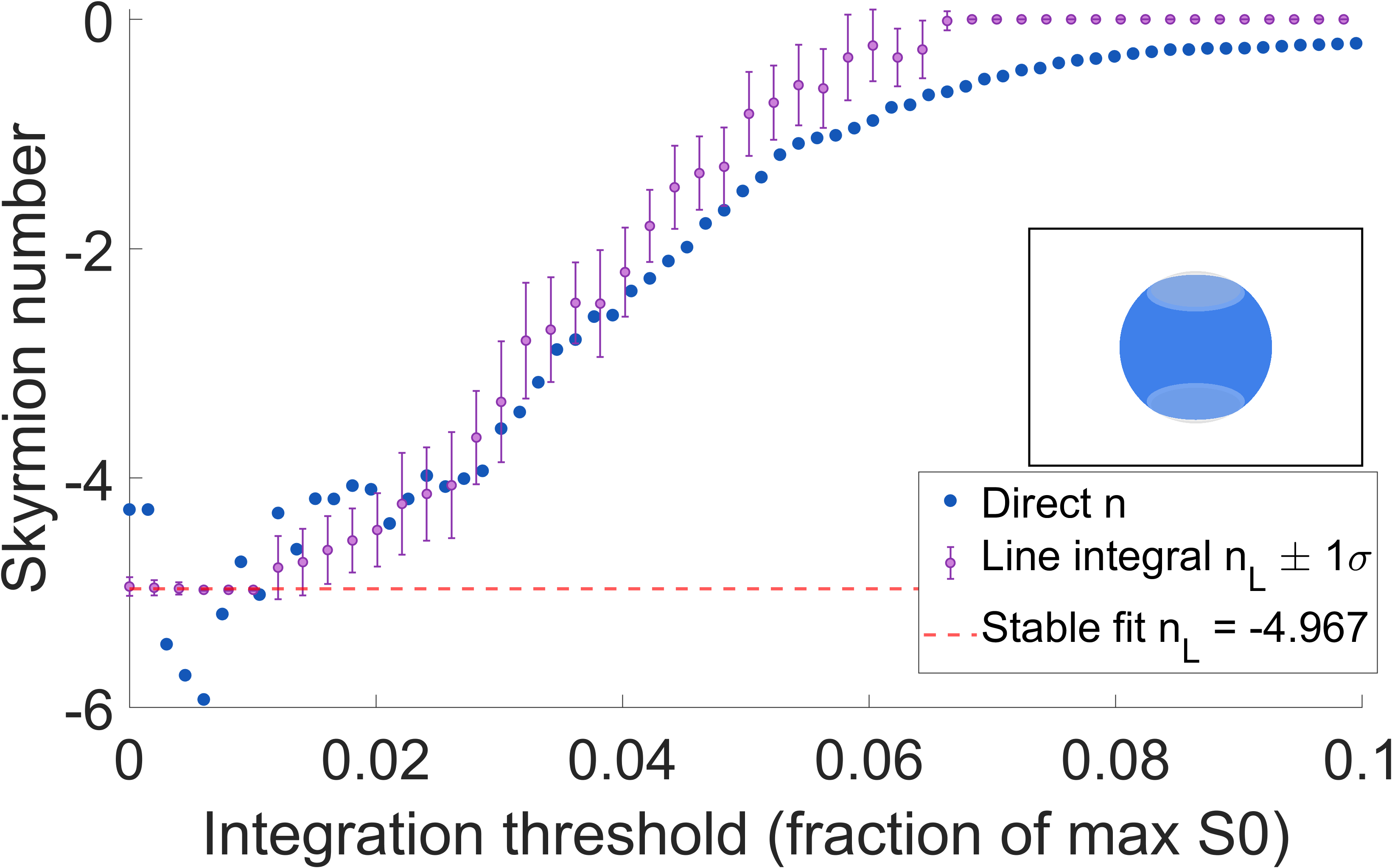}
    \end{minipage}
	\caption{Extracting topological features from derived classical Stokes parameters. The figure shows a comparison between the direct surface integral method and a contour integral approach for evaluating the Skyrmion number. The contour integral result is obtained by averaging over 400 rotation bases, selected such that singularities are confined to a high signal-to-noise region. The blue curve shows the Skyrmion number obtained via direct surface integration as a function of the intensity threshold ratio. In the inset, the blue band indicates the allowed region of rotation bases, corresponding to planes in which the polarisation phase $\Phi$ is evaluated. The purple curve shows the mean Skyrmion number obtained from the contour integrals across different threshold values, with the error bars indicating the $1\sigma$ uncertainty. The red dashed line denotes the Skyrmion number fitted from points where $\sigma = 0$, corresponding to cases in which all 400 rotation bases yield consistent values for the Skyrmion number; its exact value is provided in the legend.}
    \label{fig:supp_topologyanalysis}
\end{figure*}

\section*{Supplementary: Quantum state tomography}

 \begin{figure*}[t]
\centering\includegraphics[width=1\linewidth]{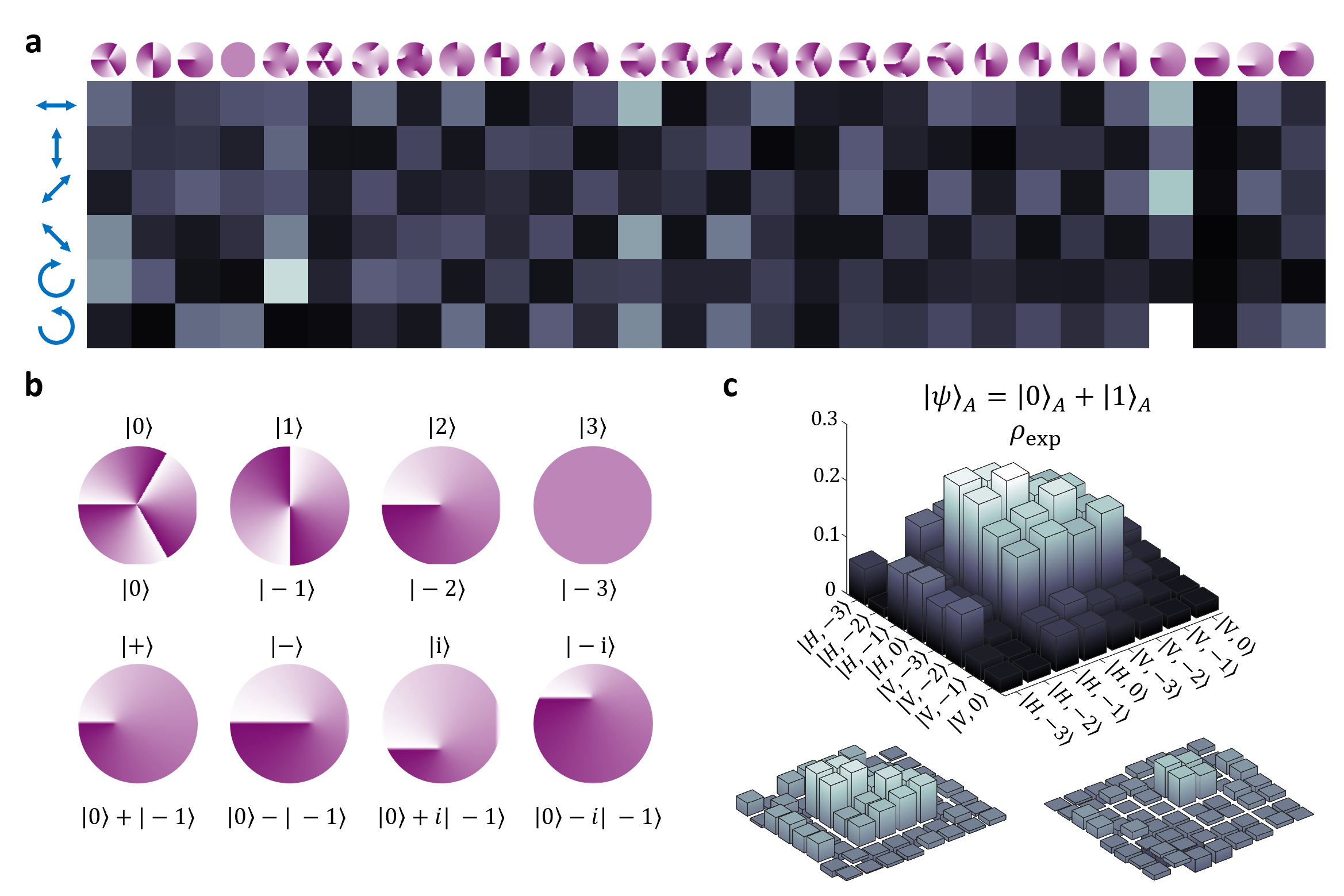}
	\caption{Quantum state tomography and density matrix reconstruction for the state given in Fig. 5 \textbf{b} of the main text. \textbf{a} Full experimental measurement matrix with polarization projections performed on photon A and columns indicating spatial projection performed on photon B with enlarged exemplar phase projections shown for the 4 basis mode projections (top row) and one example of a mutually unbiased based mode projection set (bottom row). The labels above each phase mask correspond to their computational basis representation with the labels underneath each mask indicating the OAM mode superposition. \textbf{c} Reconstructed absolute, real and imaginary components of experimentally measured density matrix.}
	\label{fig: QSTEx}
\end{figure*}

\noindent As shown in the main text, our multi-dimensional state can be decomposed into 2D hybrid states labelled by the OAM of photon A. To reconstruct the density matrix describing the resulting quantum state, we perform a quantum state tomography (QST) in the Hilbert space $\mathcal{H}=\mathcal{H}_{\mathrm{pol,A}} \otimes \mathcal{H}_{\mathrm{sp,B}}$. Here, photon~A is characterized by its polarization while photon~B is described solely by the spatial DoF, specifically OAM. The density matrix $\rho$ can be described as
\begin{equation}
    \rho =\frac{1}{2} \sum_{i,j} c_{ij} (\sigma_{i} \otimes \lambda_{j})
\end{equation}
where $d=d_\mathrm{pol,A} \times d_\mathrm{sp,B}$ is the total Hilbert space dimension, $\{\sigma_{i} \}$ denotes the Pauli operator basis for the polarization DoF of photon A and $\{\lambda_{j} \}$ are the generalised Gell-Mann operators spanning the spatial subspace of photon B, and $c_{ij}$ are the state coefficients. For the results shown in the main text, we construct an overcomplete set of measurements from tensor products of the projective measurements on each subsystem, both for the 2D hybrid states and the superimposed hybrid states. The polarization measurements for photon A uses eigenprojectors of the Pauli operators, whereas the spatial measurements on photon B uses eigenprojectors of the generalized Gell-Mann operators. Each measurement setting is represented by a projector of the form $M_k = P_a^{(A,\mathrm{pol})}\otimes P_b^{(B,\mathrm{pol})},$ and the corresponding detection probability is given by Born rule $p_k = \mathrm{Tr}(\rho M_k)$. An example of a collection of these measurement probabilities is shown in Fig.\ref{fig: QSTEx} \textbf{a} with examples of spatial mode projections given in \textbf{b}. The reconstruction of the density matrices proceeds as a linear inversion problem. After vectorizing both the density matrix and the measurement operators, the forward model takes the form $\mathbf{p} = T\,\mathrm{vec}(\rho),
$ where $T$ is the measurement matrix assembled from all projectors. A physically valid estimate of $\rho$ is obtained by minimizing the least-squared norm of the residual
\begin{equation}
  | \mathbf{p} - T\mathrm{vec}(\rho)|^{2}_{2}
\end{equation}
between the experimentally measured probabilities and the Born-rule predictions. We enforce Hermiticity, positivity, and unit trace by parametrizing the density matrix in the form $\rho = \frac{LL^{\dagger}} {\mathrm{Tr}(LL^{\dagger})},$ where $L$ is a complex lower-triangular matrix with real exponential diagonal entries. This parametrization guarantees positive semi-definiteness throughout the optimization.\\

For the analysis of the skyrmionic structure of our states, we extract the quantum Stokes parameters and determine the corresponding skyrmion numbers following the procedure outlined in Ref.~\cite{de2025revealing}.

\subsection{Entanglement witnesses}

\noindent The purity, $\gamma$, of states provides an estimate of the total signal-to-noise ratio (SNR) within a given experiment. Furthermore, the purity quantifies the degree of mixture of a state given by 
\begin{equation}
\gamma = \text{Tr} \left( \rho^2 \right) \geq \frac{1}{d}
\end{equation}
\noindent where $\gamma=1$ indicates a pure state and $\gamma = \frac{1}{d}$ indicates a maximally mixed state. For the states considered in this work, $\gamma$ scales with the dimension of the spatial qudit such that $\gamma = \frac{1}{4d_{sp,B}}$.\\

\noindent The fidelity also served as an entanglement witness for our states, used to analytically compare our measured state, $\rho$ against the closest, pure maximally entangled state $\rho_T=|\Psi\rangle\langle\Psi|$,
\begin{equation}
    F =\left( \text{Tr}  \left( \sqrt{  \sqrt{\rho_T}\rho \sqrt{\rho_T}  }  \right) \right)^2.
\end{equation}
The fidelity is 0 if the states are orthogonal or 1 when they are identical up to a global phase.\\

For the 2D hybrid states, the concurrence serves as a measure of the degree of entanglement between photon A and B, described by
\begin{equation}
    C(\rho) = \text{max} \{ 0, \lambda_1 -\lambda_2- \lambda_3 - \lambda_4 \},
\end{equation}
where $\lambda_i$ are eigenvalues of the operator $R = \text{Tr} \left( \sqrt{  \sqrt{\rho} \tilde{\rho} \sqrt{\rho}  }  \right)$ in descending order and $\tilde{\rho} = \sigma_{y} \otimes \sigma_{y} \rho^* \sigma_{y} \otimes \sigma_{y}$ with $\sigma_y = \begin{bmatrix}
  0 & -i\\ i & 0
\end{bmatrix}.$ The concurrence ranges from 0 for separable states to 1 for maximally entangled states.\\

\begin{equation}
    C = \text{max}\{0,{\lambda_1} - \sum_{i=2}{\lambda_i}\},
\end{equation}

where $\lambda_i$ are the eigenvalues of the operator $R = \sqrt{\sqrt{\rho}\tilde{\rho}} \sqrt{\rho}$, in descending order, with $\tilde{\rho} = \sigma_y \otimes \sigma_y\rho^{*}\sigma_y \otimes \sigma_y$. The concurrence ranges from 0 to 1, where a value of zero implies a separable state whereas a value of 1 indicates a maximally entangled state.


\end{widetext}

\end{document}